\documentclass[a4paper,oldfontcommands,12pt]{memoir}
\usepackage{textcomp}
\usepackage{sty/aas_macros}

\usepackage[a4paper,margin=1in]{geometry}
\usepackage{amsthm}
\usepackage{amsmath}
\usepackage{amsfonts}
\usepackage{amssymb}
\usepackage{scrpage2}	
\usepackage{color}
\usepackage{float}

\usepackage{dsfont}
\usepackage{etoolbox}
\usepackage{sty/subfigure}

\usepackage{graphicx}

\usepackage{qcircuit}

\usepackage{hyperref}

\usepackage{cleveref}

\usepackage{cite}
\usepackage{longtable}

\theoremstyle{plain}

\let\bbordermatrix\bordermatrix
\patchcmd{\bbordermatrix}{8.75}{4.75}{}{}
\patchcmd{\bbordermatrix}{\left(}{\left[}{}{}
\patchcmd{\bbordermatrix}{\right)}{\right]}{}{}

\DeclareMathOperator{\D}{d\!} 

\newcommand{\bra}[1]{\left\langle #1\right|}
\newcommand{\ket}[1]{\left|#1\right\rangle}
\newcommand{\braket}[2]{\langle #1|#2\rangle}

\crefname{table}{Tab.}{Tabs.}
\crefname{section}{Sec.}{Secs.}
\Crefname{section}{Section}{Sections}
\crefname{figure}{Fig.}{Figs.}
\Crefname{figure}{Figure}{Figures}
\crefname{equation}{Eq.}{Eqs.}
\Crefname{equation}{Equation}{Equations}
\crefname{pluralequation}{Eqs.}{Eqs.}
\creflabelformat{pluralequation}{(#2#1#3)}
\crefname{chapter}{Chap.}{Chaps.}
\Crefname{chapter}{Chapter}{Chapters}

\chapterstyle{madsen}
\setsecnumdepth{subsection}

\captionnamefont{\sffamily}
\captiontitlefont{\small}

\begin{document}


\pagestyle{empty}

\begin{center}
\LARGE{\textsc{Masterarbeit in Physik}}\\
\vspace{0.5cm}

\underline{\hspace{15 cm}}
\vspace{0.1 cm}\\
\huge{\textbf{Quantum Error Correction with the GKP Code and Concatenation with Stabilizer Codes}}\\
\vspace{0.1 cm}
\underline{\hspace{15 cm}}
\vspace{2cm}\\

\Large{von}\\
\vspace{0.5cm}

\LARGE{\textbf{Yang Wang}}
\vspace{1.8 cm}\\

\Large{Rheinisch-Westf\"alisch Technische Hochschule Aachen \\
Fakult\"at f\"ur Mathematik, Informatik und Naturwissenschaften} \\
\vspace{2 cm}

\Large{07.\ Decmber 2017}\\
\vspace{2cm}

\begin{Large}

angefertigt am\\
\vspace{0.5cm}

JARA-Institut f\"ur Quanteninformation
\vspace{1.5cm}

vorgelegt bei\\
\vspace{0.5cm}

Prof.\ Dr.\ rer.\ nat.\ Barbara M.\ Terhal

\end{Large}

\end{center}


\newpage
\phantom{hallo}
\newpage
\pagestyle{empty}

\begin{center}

\phantom{hallo}
\vspace{5cm}

\large{\textbf{Eidesstattliche Erkl\"arung}}
\end{center}
{\setlength{\parindent}{0mm}
\vspace{0.8cm}

Hiermit erkl\"are ich, Yang Wang, an Eides statt, dass ich die vorliegende Masterarbeit mit dem Titel \textit{``Quantum Error Correction with the GKP code and Concatenatino with Stabilizer Codes ''} selbstst\"andig und ohne unzul\"assige fremde Hilfe verfasst habe. Ich habe keine anderen als die angegebenen Quellen und Hilfsmittel benutzt sowie w\"ortliche und sinngem\"a{\ss}e Zitate kenntlich gemacht. Die Arbeit hat in gleicher oder \"ahnlicher Form noch keiner Pr\"ufungsbeh\"orde vorgelegen.

\vspace{1.5cm}

Aachen, den 07.\ December 2017

\vspace{1cm}

\underline{\hspace{4 cm}}\vspace{0.03 cm}\\
(Yang Wang)}


\newpage
\begin{center}
\LARGE{\textbf{Acknowledgement}}
\end{center}
\vspace{0.3cm}

\noindent First and foremost, I would like to express my gratitude to my supervisor Barbara Terhal for her guidance, patience and for many insightful discussions. She is so experienced and her knowledge in this field made my thesis possible. It's really good to have the opportunity to work under her supervision.

I shall extend my thanks to Daniel Weigand and Kasper Duivenvoorden for their useful discussions as well as helping me with many technical things. There are also thanks to many friends in Aachen, especially Huanbo Sun, for their support when I was new in Germany.

Finally, I want to thank my family. My parents funded my whole master study and always encourage me when I feel down. Also my boyfriend, Zhichuan Liang for his love and patience all the time.

\newpage
\phantom{hallo}
\pagestyle{plain}
\newpage
\pagenumbering{roman}

\begin{center}
\large{\textbf{Abstract}}
\end{center}
\vspace{0.3cm}

\noindent Gottesman, Kitaev and Preskill have proposed a scheme to encode a qubit in a harmonic oscillator \cite{gottesman2001encoding}, which is called the GKP code. It is designed to be resistant to small shift errors contained in momentum and position quadratures. Thus there's some intrinsic fault tolerance of the GKP code.

In this thesis,we propose a method to utilize all the information contained in the continuous shifts, not just simply map a GKP-encoded qubit to a normal qubit. This method enables us to do maximum-likelihood decisions and thus increase fault-tolerance of the GKP code. This thesis shows that the continuous nature of the GKP code is quite useful for quantum error correction.

\addtodef{\listoffigures}{\clearpage\pagestyle{plain}}{}
\addtodef{\listoftables}{\clearpage\pagestyle{plain}}{}
\addtodef{\tableofcontents}{\clearpage\pagestyle{plain}}{}

\tableofcontents



\mainmatter

\chapter{Introduction}

At the beginning of the twenty-first century, Gottesman, Kitaev and Preskill proposed a code called the GKP code, which encodes a qubit into a harmonic oscillator \cite{gottesman2001encoding}. The GKP code is designed to be resistant to small shifts in momentum and position quadratures of a harmonic oscillator because of its continuous nature. Apart from the intrinsic fault-tolerance, Clifford gates and error correction on the GKP encoded qubits can be constructed using linear optical elements only, and universal quantum computation requires photon counting in addition. These features make the GKP code a promising encoding scheme in experiments.
 
Often, people simply map it back to a qubit with an average error rate, which means that some information has been neglected. Gottesman \textit{et al.} \cite{gottesman2001encoding} proposed an error correction scheme similar to the Steane method of quantum error correction \cite{steane1996error}, which efficiently corrects small shift errors with some error rate. For repeated Steane error corrections, Glancy and Knill found a $\frac{\sqrt{\pi}}{6}$ error bound  \cite{glancy2006error}. However, these averaged rates are not very interesting, because there's nothing different from a normal qubit.

Thus in this thesis, we try to take the neglected information into account,  we call it the GKP error information.  With this information, decoding of various stabilizer codes concatenated with the GKP code is also analyzed. And the most significant result in this thesis is for the toric code, its error threshold can be achieved with much nosier GKP-encoded qubits.

The thesis is structured as follows: \cref{sec:background} can be regarded as the definition of the GKP code, including its encoding scheme and an analysis of its internal shift errors due to finite squeezing. \cref{sec:QC-GKP} reviews how to use the GKP code in quantum computation: Clifford gates with only linear optical elements and non-symplectic gates via photon counting. The Steane error correction scheme and how nicely it fits within the framework of cluster states are also discussed. In \cref{sec:QEC-continuous}, we start to take the neglected GKP error information into account, with which the decoding schemes of various stabilizer codes concatenated with the GKP code are modified. Finally, \cref{sec:conclusion-outlook} summarizes the results of this thesis and discusses possible future work.

\chapter{Theoretical Background}
\label{sec:background}

This chapter introduces the theoretical foundations of the GKP code. \cref{sec:encoding_oscillator} introduces how to encode a qubit into a harmonic oscillator. \cref{sec:Finitely_squeezed_gkp} introduces the concept of approximate GKP code with finite squeezing, which is still useful since the GKP code is designed to be protected against small shift errors. In \cref{sec:shifted_code_states}, we use a handy tool called the shifted code states to analyze the probability distributions of the internal errors due to finite squeezing. It will be shown that the distributions can be approximated by Gaussian distributions when code states are squeezed enough. \cref{sec:Gaussian_shift_error} considers stochastic shift errors from an error channel named "Gaussian Shift Error Channel". This section also compares stochastic errors with coherent errors due to finite squeezing.

\section{Ideal GKP Code States}
\label{sec:encoding_oscillator}

In this section, we discuss the encoding scheme of the GKP code introduced by Gottesman \textit{et al.} \cite{gottesman2001encoding}. It encodes a state of a finite dimensional quantum system in an infinite-dimensional system, i.e. encoding a qubit into a harmonic oscillator.

For clarity, we first introduce some basic notations which will be used later. For a harmonic oscillator, the position and momentum operators are defined as:
\begin{equation}
\begin{split}
\hat{q} = \frac{1}{\sqrt{2}} ( \hat{a} + \hat{a}^{\dag}),\\
\hat{p} = \frac{i}{\sqrt{2}} ( \hat{a} - \hat{a}^{\dag}),
\end{split}
\label{eq:qp_aadag}
\end{equation}
where $\hat{a}$ and $\hat{a}^{\dag}$ are creation and annihilation operators. They satisfy the canonical commutation relation:
\begin{align}
[\hat{q}, \hat{p}] = i.
\label{eq:qp_commutation}
\end{align}
Their eigenstates are called quadrature states, which are defined as $\ket{q}$ and $\ket{p}$:
\begin{equation}
\begin{split}
\hat{q} \ket{q} = q \ket{q},\\
\hat{p} \ket{p} = p \ket{p},
\end{split}
\end{equation}
where $q,p$ are arbitrary real numbers. And they are connected by the Fourier transformation relation:
\begin{align}
\ket{q} = \int \frac{\D p}{\sqrt{2\pi}}e^{ipq}\ket{p}.
\label{eq:qp_fourier}
\end{align} 
Since the shifts in both quadratures are continuous, any specific shifts $u,v$ in the $\hat{q}$ and the $\hat{p}$ quadratures can be written as displacement operators:
\begin{equation}
\begin{split}
&\ket{q+u} = e^{- i u \hat{p}} \ket{q},\\
&\ket{p+v} = e^{+ i v \hat{q}} \ket{p},\\
\end{split}
\end{equation}
where $u,v$ are arbitrary real numbers. Expand the displacement operators in Taylor series and use the commutation relation in \cref{eq:qp_commutation}, one can easily check the above relations.


\subsection{Encoding a Qubit into a Harmonic oscillator}
\label{sec:encoding scheme}
In order to introduce the GKP code, we first denote two displacement operators with displacement $2\sqrt{\pi}$ as:
\begin{equation}
\begin{split}
\hat{S}_q = e^{-i 2\sqrt{\pi} \hat{p}},\quad \hat{S}_p = e^{i2\sqrt{\pi} \hat{q}}.
\end{split}
\end{equation}
With the Baker-Campell-Hausdorff formula that $ e^A e^B = e^Be^Ae^{[A,B]}$, it's easy to check that the two displacement operators commute to each other:
\begin{equation*}
[\hat{S}_q, \hat{S}_p] = 0.
\end{equation*}
Thus we call $\hat{S}_q$ and $\hat{S}_p$ as stabilizer operators because we can find simultaneous eigenstates of them. It's clear that both momentum $p$ and position $q$ of a simultaneous eigenstate are sharply determined, i.e.
\begin{align*}
q ,p = 0 \mod \sqrt{\pi} = n \sqrt{\pi},\quad n \in \mathbb{Z}.
\end{align*}
This subspace stabilized by the stabilizer operators is two-dimensional. To see this, we define the logical $\ket{\bar{0}}$ and $\ket{\bar{1}}$ expanded in the quadrature states $\ket{q}$ as:
\begin{equation}
\begin{split}
& \ket{\bar{0}} = \sum_n \delta(q - 2 n \sqrt{\pi})\ket{q}= \sum_n \ket{q = 2 n \sqrt{\pi}} ,\\
& \ket{\bar{1}} = \sum_n \delta(q - (2 n +1) \sqrt{\pi})\ket{q}= \sum_n \ket{q = (2 n +1) \sqrt{\pi}}.
\end{split}
\label{eq:01_q_quadrature}
\end{equation}
It's easy to see that $\ket{\bar{0}}$ and $\ket{\bar{1}}$ are connected by a displacement operator with displacement $\sqrt{\pi}$, we define this displacement operator as the logical $\overline{X}$ operator:
\begin{align*}
\overline{X} = e^{-i\sqrt{\pi}\hat{p}}.
\end{align*}
It's easy to check that $\overline{X}$ commutes with the stabilizer operators and $\overline{X}^2 = \hat{S}_q$. Symmetrically, logical $\ket{\overline{+}}$ and logical $\ket{\overline{-}}$ are defined as:
\begin{equation}
\begin{split}
& \ket{\bar{+}} = \sum_n \delta(p - 2 n \sqrt{\pi})\ket{p}= \sum_n \ket{p = 2 n \sqrt{\pi}} ,\\
& \ket{\bar{-}} = \sum_n \delta(p - (2 n +1) \sqrt{\pi})\ket{p}= \sum_n \ket{p = (2 n +1) \sqrt{\pi}},
\end{split}
\label{eq:+-_q_quadrature}
\end{equation}
and the corresponding logical operator is defined as:
\begin{align*}
\overline{Z} = e^{+i\sqrt{\pi}\hat{q}} = \hat{S}_p^{\frac{1}{2}}.
\end{align*}
Armed with the Fourier transformation relation in \cref{eq:qp_fourier} and the Poisson summation rule $\sum\limits_{n=-\infty}^{\infty} \delta(x-nT) = \sum\limits_{n=-\infty}^{\infty}\frac{1}{T} e^{i2\pi\frac{k}{T}x}$, it's easy to check that:
\begin{equation}
\begin{split}
\ket{\bar{0}} & =\sum_n \delta(q - 2 n \sqrt{\pi})\ket{q} =\int \frac{\D p}{\sqrt{2\pi}} \sum_n e^{ip2n\sqrt{\pi}}\ket{p}\\
              & = \frac{1}{\sqrt{2}} \sum_n \delta(p - n \sqrt{\pi})\ket{p}\\
              & = \frac{1}{\sqrt{2}}( \ket{\overline{+}} + \ket{\overline{-}}). 
\end{split}
\label{eq:0_p_quadrature}
\end{equation}
Similarly we can also transform other states:
\begin{equation*}
\begin{split}
\ket{\bar{1}} &  = \frac{1}{\sqrt{2}}( \ket{\overline{+}} + \ket{\overline{-}}) \propto  \sum_n (-1)^n \delta(p - n \sqrt{\pi})\ket{p}\\
\ket{\bar{+}} & = \frac{1}{\sqrt{2}}( \ket{\overline{+}} + \ket{\overline{-}}) \propto \sum_n  \delta(q - n \sqrt{\pi})\ket{q}\\
\ket{\bar{-}} & = \frac{1}{\sqrt{2}}( \ket{\overline{+}} + \ket{\overline{-}}) \propto \sum_n (-1)^n \delta(q - n \sqrt{\pi})\ket{q}.\\
\end{split}
\end{equation*}
Thus the ideal GKP code is defined and a qubit is encoded into a harmonic oscillator. Here the quadrature $\hat{q}$ and the quadrature $\hat{p}$ are symmetric, but it's not necessary to be like this, one can find a more general definition\cite{gottesman2001encoding} of the GKP code in the paper of Gottesman \textit{et al.}

\subsection{Resistance to Small Shift Errors}
\label{sec:intrinsic_fault_tolerant}

The GKP code is designed to be resistant to small shift errors, small shift errors can be detected and then be corrected. Imagine we prepared an arbitrary ideal GKP state $\ket{\overline{\psi}} = \alpha \ket{\overline{0}} + \beta \ket{\overline{1}}$, which is stabilized by $\hat{S}_q$ with eigenvalue $+1$. However, if this state is subjected to some small shift error $u$ in the $\hat{q}$ quadrature, the eigenvalue of $\hat{S}_q$ will no longer be exactly $+1$ (the analysis for shift error in the $\hat{p}$ quadrature and $\hat{S}_p$ would be the same):
\begin{align*}
S_{\hat{q}} e^{-i u\hat{p}} \ket{\bar{\psi}} & = e^{i 2\sqrt{\pi}\hat{q}} e^{-i u\hat{p}} \ket{\bar{\psi}} = e^{-i u\hat{p}} e^{i 2\sqrt{\pi}\hat{q}} e^{2\sqrt{\pi}u[\hat{q},\hat{p}]}\ket{\bar{\psi}}\\
&=e^{i 2\sqrt{\pi}u } e^{-i u\hat{p}} \ket{\bar{\psi}}
\end{align*}
where we use the Baker-Campell-Hausdorff formula that $ e^A e^B = e^Be^Ae^{[A,B]}$. Thus the eigenvalue has a phase $2\sqrt{\pi}u$, we can obtain some information about it since we assume small error. If we could do good estimation of this phase, we can just shift the code state back and thus correct the error.

\begin{figure}
\centering
\begin{minipage}{\textwidth}
\[
\Qcircuit @C=1.2em @R=2em {
\lstick{ \ket{\tilde{\psi} } } &\qw & \gate{\hat{S}_q} &\qw  \\
\lstick{ \ket{+} } &\qw & \ctrl{-1}               &\measure{Measure}
}\]
\end{minipage}
\caption{Circuit of Controlled-$\hat{S}_q$ gate. where $\ket{ \tilde{\psi} }$ is an ideal GKP code state subjected to shift error in the $\hat{q}$ quadrature. The controlled-$\hat{S}_q$ consists of two CNOT gates, it is controlled by the ancilla qubit in state $\ket{+}$. Measure the ancilla qubit in basis $\{\ket{+},\ket{-} \}$. }
\label{fig:controlled_sq}
\end{figure}
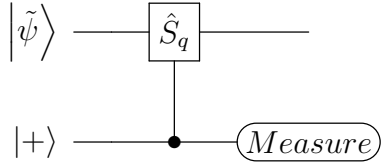

We can measure the phase by applying a controlled $\hat{S}_q$  on the qubit, see \cref{fig:controlled_sq}. This controlled $\hat{S}_q$ gate consists of two CNOT gate controlled by the ancilla prepared in $\ket{+}$. After measuring the ancilla in the basis $\{\ket{+},\ket{-}\}$, we can obtain information about the phase that is because the CNOT gate moves the shift error in the $\hat{q}$ quadrature from the data qubit $\ket{\tilde{\psi}}$ to the ancilla, see details in \cref{sec:error propagation}. 

But the phase estimation protocol uses bare physical qubits to control the displacement and requires many rounds of measurements, which makes it not fault-tolerant for quantum error correction\cite{terhal2016encoding}\cite{WeigandMaster2015}. On the other hand, if we have access to GKP-encoded ancilla qubits, we do Steane error correction instead (See details in \cref{sec:steane_qec} and \cref{sec:further_steane}), which requires only one CNOT gate. More importantly, Steane error correction is fault-tolerant for quantum error correction, see details in \cref{sec:fault-tolerant_Steane}.

\section{Finitely Squeezed States}
\label{sec:Finitely_squeezed_gkp}

\begin{figure}
\begin{minipage}[t]{0.5\textwidth}
\centering
\includegraphics[width=1\textwidth]{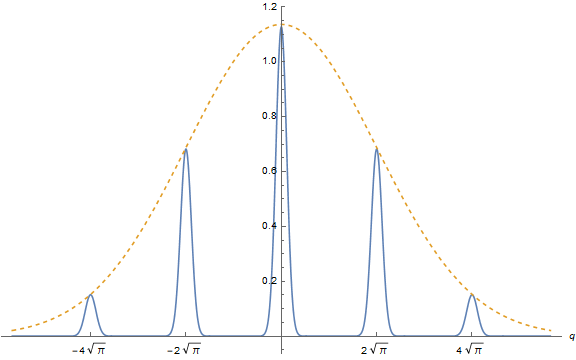}
\end{minipage}
\begin{minipage}[t]{0.5\textwidth}
\centering
\includegraphics[width=1\textwidth]{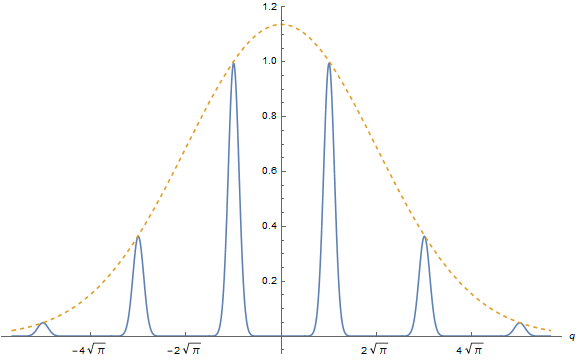}
\end{minipage}
\caption{Blue lines represent the absolute value of the wave functions $\vert \Psi(q) \vert^2$ of approximate code states in the $\hat{q}$ quadrature with $\Delta=0.2$. Left figure is for the approximate state $\ket{\tilde{0}}$. Right one is for the approximate state $\ket{\tilde{1}}$. The dashed orange line is the Gaussian envelope.}
\label{fig:approx_state}
\end{figure}

Strictly speaking, the code words defined in \cref{sec:encoding_oscillator} are not physical. These ideal states are not normalizable and infinite squeezing requires infinite energy. However, the GKP code is designed to be protected from small shift errors as discussed in \cref{sec:intrinsic_fault_tolerant}, thus some small spreads in both quadratures are allowed. 

It's natural to replace $\delta$ functions in ideal states by normalized Gaussian functions with variance $\Delta$. In order to make it symmetric in both quadratures, these Gaussian functions are weighted by a Gaussian envelope \cite{gottesman2001encoding}. The wave functions of these approximate code states are defined as (up to normalization):
\begin{equation}
\begin{split}
&\ket{\tilde{0}} \propto \sum_{s=-\infty}^{\infty}\int_{-\infty}^\infty e^{-\frac{\Delta^2}{2}(2s)^2\pi}e^{-\frac{1}{2\Delta^2}(q-2s\sqrt{\pi})^2}\ket{q} \D q,  \\
&\ket{\tilde{1}} \propto \sum_{s=-\infty}^{\infty}\int_{-\infty}^\infty e^{-\frac{\Delta^2}{2}(2s+1)^2\pi}e^{-\frac{1}{2\Delta^2}(q-(2s+1)\sqrt{\pi})^2} \ket{q} \D q .
\end{split}
\label{eq:approximate_states}
\end{equation}
The absolute value of the wave functions of $\ket{\tilde{0}}$ and $\ket{\tilde{1}}$ in the $\hat{q}$ quadrature are shown in \cref{fig:approx_state}, where $\Delta =0.2$. Similarly we have  $\ket{\tilde{+}}$ and $\ket{\tilde{-}}$ expanded in the $\hat{p}$ quadrature written as:
\begin{equation}
\begin{split}
&\ket{\tilde{+}} \propto \sum_{s=-\infty}^{\infty}\int_{-\infty}^\infty e^{-\frac{\Delta^2}{2}(2s)^2\pi}e^{-\frac{1}{2\Delta^2}(p-2s\sqrt{\pi})^2}\ket{p} \D p,  \\
&\ket{\tilde{-}} \propto \sum_{s=-\infty}^{\infty}\int_{-\infty}^\infty e^{-\frac{\Delta^2}{2}(2s+1)^2\pi}e^{-\frac{1}{2\Delta^2}(p-(2s+1)\sqrt{\pi})^2} \ket{p} \D p.
\end{split}
\end{equation}
Similarly With the Fourier transformation relation in \cref{eq:qp_fourier} and the Poisson summation rule $(a)^{-1/2} \sum_{s=-\infty}^{\infty} e^{-\pi s^2/a} e^{2\mathrm{i}\pi t b} = \sum_{m=-\infty}^{\infty} e^{-\pi a (m-b)^2}$\cite{terhal2016encoding}, it's easy to check that these states have the correct form in the conjugate quadratures, here we take $\ket{\tilde{0}}$ for example:
\begin{align*}
\ket{\tilde{0}} & = \sum_{s=-\infty}^{\infty}\int_{-\infty}^\infty e^{-\frac{\Delta^2}{2}(2s)^2\pi}e^{-\frac{1}{2\Delta^2}(q-2s\sqrt{\pi})^2}\ket{q} \D q\\
&= \sum_{s=-\infty}^{\infty}\int_{-\infty}^\infty e^{-\frac{\Delta^2}{2}(2s)^2\pi}e^{-\frac{1}{2\Delta^2}(q-2s\sqrt{\pi})^2}\D q \int \frac{\D p}{\sqrt{2\pi}}e^{\mathrm{i}pq}\ket{p}\\
& = \frac{\Delta}{\sqrt{2}} e^{-\frac{p^2\Delta^2}{2}} \cdot \underbrace{\int\limits_{-\infty}^{\infty} e^{-\frac{1}{2\Delta^2}(q-2s\Delta^2\sqrt{\pi}-i\Delta^2 p)^2} \D q }_{\text{Gaussian Integral}}  \cdot \underbrace{ \frac{\sqrt{2}}{\Delta} \sum\limits_{s=-\infty}^{\infty} e^{-\frac{2\pi s^2}{\Delta^2}} e^{2i\pi \Delta^2\frac{p}{\sqrt{\pi}} } }_{\text{Poisson Summation}} \\
&\propto \sum_{s=-\infty}^{\infty}\int_{-\infty}^\infty e^{-\frac{\Delta^2}{2}s^2\pi}e^{-\frac{1}{2\Delta^2}(p-s\sqrt{\pi})^2}\ket{p} \D p
\end{align*}
where we take the approximation that $ p \approx s \sqrt{\pi}$ in the last step.
For an arbitrary encoded qubit state $\ket{\tilde{\Psi}}$, we can write it as an ideal state $\ket{\bar{\Psi}}$ subjected to shift errors (up to normalization and phase factors):
\begin{align}
\ket{\tilde{\Psi}} \propto \int_{-\infty}^\infty \int_{-\infty}^\infty e^{-\frac{1}{2}\left(\frac{u^2}{\Delta^2}\right)} e^{-iu\hat{p}+iv\hat{q}} \ket{\bar{\Psi}} \D u \D v,
\label{eq:approximate_wave_function}
\end{align}
As seen in \cref{eq:approximate_wave_function}, approximate code states $\ket{\tilde{0}}$ and $\ket{\tilde{1}}$ are only approximately orthogonal, even with perfect measurements of the qubit, it is still possible to obtain logical errors. However, as discussed in \cref{sec:intrinsic_fault_tolerant}, GKP code states subjected to small shift errors is still useful, and it will be shown that the Steane error correction 
is fault-tolerant, see \cref{sec:fault-tolerant_Steane}.

Similar to the ideal code states, the wave functions of the approximate code states remain symmetric with respect to 0 in both quadratures, and when the squeezing parameter goes to zero, the approximate code states are reduced back to ideal states.

\section{Internal Shift Errors due to Finite Squeezing}
\label{sec:shifted_code_states}

The GKP code is designed to be resistant to small shift errors, thus the finitely squeezed GKP code is still useful. In this section we use a practical tool, the shifted code states\cite{glancy2006error}, to analyze the probability distributions of the shift errors in the momentum and the position quadratures. It will be shown that for highly squeezed GKP code states, the probability distributions in both quadratures are independent Gaussian distributions, and the variance is related to the squeezing parameter $\Delta$.

\subsection{Shifted Code States in Two Quadratures}

For simplicity we assume that the approximate code is squeezed enough, the shift error $u$ in the $\hat{q}$ quadrature and $v$ in the $\hat{p}$ quadrature are localized enough that $ \vert u \vert, \vert v \vert  \leq \sqrt{\pi}$. The original definition of the shifted code states is\cite{glancy2006error}:
\begin{align}
\ket{u,v} &= \frac{1}{\sqrt[4]{\pi}} e^{-i v \hat{q}} e^{-i u \hat{p}}\ket{\bar{0}},
\label{eq:shifted_code_state}
\end{align}
In \cref{eq:01_q_quadrature} and \cref{eq:0_p_quadrature}, we see that $\ket{\overline{0}}$ is a superposition of delta functions in the $\hat{q}$ and also the $\hat{p}$ quadratures. Any neighboring peaks in the $\hat{q}$ quadrature differ by a shift equal to $2\sqrt{\pi}$, while there's only $\sqrt{\pi}$ in the $\hat{p}$ quadrature. Thus according to this periodicity, the $u,v$ are restricted that:
\begin{align*}
u \in [-\sqrt{\pi}, \sqrt{\pi}], \quad v \in [-\sqrt{\pi}/2, \sqrt{\pi}/2].
\end{align*}
It's easy to check that the shifted code states are orthogonal:
\begin{align*}
\braket{u,v}{u',v'} &= \frac{1}{\sqrt{\pi}} \sum_{s,s'} e^{i (v (2s\sqrt{\pi}+u) - v'(2  s' \sqrt{\pi}+u'))}\underbrace{\braket{2 s \sqrt{\pi}+u}{2 s' \sqrt{\pi}+u'}}_{\delta_{s,s'} \delta(u-u')} \\
&= \frac{1}{\sqrt{\pi}}   e^{i u (v - v')}\sum_{s} e^{i s 2 \sqrt{\pi} (v - v')}\delta(u-u')\\
&=\delta(v-v')\delta(u-u'),
\end{align*}
where $\delta(x) = \frac{1}{2\pi}\sum_s e^{isx}$ and $\delta(ax) = \frac{1}{|a|}\delta(x)$. On the other hand, the shifted code states can represent an arbitrary state shifted from $\ket{\overline{0}}$ and $\ket{\overline{1}}$, thus we can expand an arbitrary state $\ket{\tilde{\phi}}$ in the shifted code states, which means that the shifted code states form a complete basis\cite{glancy2006error}\cite{WeigandMaster2015}.

However, there's a problem when we want to use $\ket{u,v}$ to analyze the shift error in the $\hat{p}$ quadrature.
For arbitrary two states $\ket{u,v_1}$ and $\ket{u,v_2}$, where $\vert v_1 - v_2 \vert = \sqrt{\pi}$, it's completely impossible to tell the difference between them (up to some phase factors):
\begin{align*}
\ket{u,v_1} = \ket{u,v_2 \pm \sqrt{\pi}}_q = \frac{1}{\sqrt[4]{\pi}} e^{-i v_2 \hat{q}} e^{\pm i \sqrt{\pi} \hat{q}} e^{-i u \hat{p}}\ket{\bar{0}} = \ket{u,v_2}, 
\end{align*}
which means that we cannot use $\ket{u,v}$ defined above to analyze the shift error in the $\hat{p}$ quadrature where the error is not restricted in $[-\sqrt{\pi}/2, \sqrt{\pi}/2]$. Since we can use it to analyze the shift error in the $\hat{q}$ quadrature, we redefine it as the shifted code states in the $\hat{q}$ quadrature with subscript $q$ as:
\begin{equation}
\begin{split}
\ket{u,v}_q &= \ket{u,v}.
\end{split}
\end{equation}
In order to analyze the shift error in the $\hat{p}$ quadrature, it's natural to replace the $\ket{\overline{0}}$ by $\ket{\overline{+}}$, thus the so-called shifted code states in the $\hat{p}$ quadrature is defined as:
\begin{equation}
\begin{split}
\ket{u,v}_p &= \frac{1}{\sqrt[4]{\pi}} e^{-i v \hat{q}} e^{-i u \hat{p}}\ket{\bar{+}},\\
u & \in [-\sqrt{\pi}/2,\sqrt{\pi}/2],\\
v & \in [-\sqrt{\pi},\sqrt{\pi}].
\end{split}
\end{equation}
In conclusion, we use $\ket{u,v}_p$ to analyze the shift error in the $\hat{p}$ quadrature and $\ket{u,v}_q$ in the $\hat{q}$ quadrature, given that all shift errors are in the range $[-\sqrt{\pi},\sqrt{\pi}]$.

The above analysis explains why the periodicity of $u,v$ for the original definition is asymmetric. The shifted code states are non-physical states just like the ideal code states.  But a physical state can be expressed as a superposition of shifted code states. In next section, we introduce how to use the shifted code states to analyze the probability distributions of the shift errors.




\subsection{Approximate Internal Shift Error Probability Distributions}
\label{sec:uv_wavefunction}

As discussed above, the shifted code states form a complete orthonormal basis,  an arbitrary state $\ket{\Phi}$ can be expanded in this basis:
\begin{align}
\ket{\Phi} = \int\limits_{-\sqrt{\pi}}^{\sqrt{\pi}} \D u
\int\limits_{-\frac{\sqrt{\pi}}{2} }^{\frac{\sqrt{\pi}}{2}} \D v \braket{u,v}{\Phi} \cdot \ket{u,v},
\end{align}
where we use $\ket{u,v} = \ket{u,v}_q$ to get the probability distribution of shift error $u$ in the $\hat{q}$ quadrature, the analysis with $\ket{u,v}_p$ for shift error in the $\hat{p}$ quadrature would be the same. 

\begin{figure}
\begin{minipage}[t]{0.5\textwidth}
\centering
\includegraphics[width=1\textwidth]{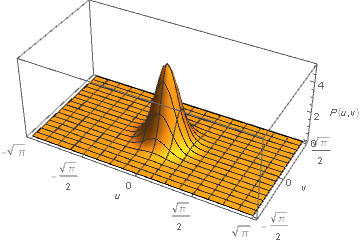}
\end{minipage}
\begin{minipage}[t]{0.5\textwidth}
\centering
\includegraphics[width=1\textwidth]{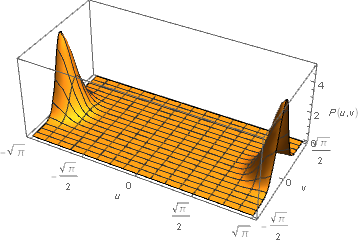}
\end{minipage}
\caption{Probability density of approximate code states' shift errors with the squeezing parameter $\Delta = 0.25$, using the shifted code states in the $\hat{q}$ quadrature $\ket{u,v}_q$. Where $u$, $v$ are the shift errors in the $\hat{q}$ quadrature and the $\hat{p}$ quadrature respectively. Left one is for approximate state $\ket{\tilde{0}}$. The right one is for approximate state $\ket{\tilde{1}}$. The probability distribution is periodic with respect to $u \in [-\sqrt{\pi},\sqrt{\pi}]$ and $v \in [-\sqrt{\pi}/2,\sqrt{\pi}/2]$, so only one peak in that range is shown here.}
\label{fig:prob_density_uv}
\end{figure}

Since the shifted code states are defined to represent each a specific error in the $\hat{q}$ quadrature and the $\hat{p}$ quadrature respectively. Thus the probability distribution of these errors for some state $\ket{\Phi}$ is simply given by:
\begin{align*}
P(u,v) = P(u,v|\Phi)  = |\braket{u,v}{\Phi}|^2.
\end{align*}
This distribution can be used to analyze the logical error rate of an arbitrary GKP code state, which is the main reason why Glancy and Knill introduced the shifted code states. Here we first set $\ket{\Phi}= \ket{\tilde{0}}$ :
\begin{align*}
P(u,v)=&\vert \langle u,v \vert \tilde{0} \rangle \vert^2\\ 
=&\frac{2}{\pi} \sum_{t,t'} \sum_{s,s'}e^{iv(2t-2t')\sqrt{\pi}} e^{-2\pi\Delta^2(s^2+s'^2)} e^{-\left( u+\left(2t'-2s'\right)\sqrt{\pi}\right)^2/(2\Delta^2)} e^{-\left( u+\left(2t-2s\right)\sqrt{\pi}\right)^2/(2\Delta^2)}\\
=&\frac{2}{\pi} \sum_{t,t'} \sum_{s,s'}e^{iv(2t-2t')\sqrt{\pi}} e^{-2\pi\Delta^2(s^2+s'^2)} e^{-\frac{u^2}{\Delta^2}}e^{-\frac{2\left(t+t'-s-s'\right)\sqrt{\pi} }{\Delta^2} } e^{-\frac{2\left((t-s)^2+(t'-s')^2\right)\pi}{\Delta^2}},
\end{align*}
we assume $e^{-\frac{2\pi}{\Delta^2}} \ll 1$, i.e. $\Delta \ll \sqrt{2\pi}\approx 2.5$. Thus for small squeezing parameter $\Delta$, it's safe to keep only the terms with $t = s$ and $t' = s'$, since the terms with $t\neq s$ or $t' \neq s'$ goes to 0:
\begin{equation}
\begin{split}
P(u,v) \approx &\frac{2}{\pi} \sum_{s,s'}e^{\mathrm{i} v(2s-2s')\sqrt{\pi}} e^{-2\pi\Delta^2(s^2+s'^2)} e^{-\frac{u^2}{\Delta^2}}\\
=&\underbrace{\left[\frac{2\Delta}{\sqrt{\pi}} \sum_{s,s'}e^{\mathrm{i} v(2s-2s')\sqrt{\pi}} e^{-2\pi\Delta^2(s^2+s'^2)}  \right]}_{\text{only depends on $v$}} \underbrace{\frac{1}{\frac{\Delta}{\sqrt{2}}\cdot\sqrt{2\pi}} e^{-\frac{u^2}{2\left(\frac{\Delta}{\sqrt{2}}\right)^2}} }_{\text{Gaussian only dependes on $u$} }
\end{split}
\end{equation}
Note that $P(u,v)$ is now separated into two parts, one part only depends on $u$ and the other only depends on $v$. Clearly the probability distributions of $u$ and $v$ are independent when $\Delta$ is small. One could simply integrate out $v$ over $[-\sqrt{\pi}/2,\sqrt{\pi}/2]$ to get the probability distribution of $u$:
\begin{align*}
P(u)=& \int_{-\frac{\sqrt{\pi}}{2}}^{\frac{\sqrt{\pi}}{2}}\mathrm{d}v P(u,v)\\
=&\sum_{s,s'}\underbrace{\int_{-\frac{\sqrt{\pi}}{2}}^{\frac{\sqrt{\pi}}{2}}\mathrm{d}ve^{\mathrm{i} v(2s-2s')\sqrt{\pi}}}_{\sqrt{\pi}\delta_{s,s'}} \left[\frac{2\Delta}{\sqrt{\pi}} e^{-2\pi\Delta^2(s^2+s'^2)}  \right] \underbrace{\frac{1}{\frac{\Delta}{\sqrt{2}}\cdot\sqrt{2\pi}} e^{-\frac{u^2}{2\left(\frac{\Delta}{\sqrt{2}}\right)^2}} }_{Gaussian}\\
=& \underbrace{\left[2\Delta \sum_{s}e^{-4\pi\Delta^2 s^2}  \right]}_{\text{a constant}} \underbrace{\frac{1}{\frac{\Delta}{\sqrt{2}}\cdot\sqrt{2\pi}} e^{-\frac{u^2}{2\left(\frac{\Delta}{\sqrt{2}}\right)^2}} }_{Gaussian}.
\end{align*}
with $\Delta \lesssim 0.4$, it's easy to check that the constant $2\Delta \sum_{s}e^{-4\pi\Delta^2 s^2} \to 1$. Now we proved that $u$ satisfy a Gaussian distribution with variance $\Delta/\sqrt{2}$:
\begin{align}
u\sim \mathcal{N}\left(0,\frac{\Delta}{\sqrt{2}}\right).
\label{eq:approx_gaussian}
\end{align}
The state can be written as $\ket{\Phi} = \alpha \ket{\tilde{0}}+ \beta \ket{\tilde{1}}$, $ |\braket{u,v}{\tilde{0}}|^2$ is the same as $|\braket{u,v}{\tilde{1}}|^2$ up to a shift of $\sqrt{\pi}$ in the $\hat{q}$ quadrature, see Fig.(\ref{fig:prob_density_uv}). Considering that the ideal state $\ket{\overline{1}}$ itself contains a logical shift equal to $\sqrt{\pi}$ in the $\hat{q}$ quadrature, the shift error $u$ of an arbitrary state $\ket{\Phi}$ just satisfies this Gaussian distribution as in \cref{eq:approx_gaussian}.

For the probability distribution of the shift error $v$ in the $\hat{p}$ quadrature, we expand a state as $\ket{\Phi} = \alpha \ket{\tilde{+}}+ \beta \ket{\tilde{-}}$ and use the shifted code states $\ket{u,v}_p$ to do the same analysis, it's easy to find that $v$ also satisfies the same Gaussian distribution as in \cref{eq:approx_gaussian}. Here $u,v$ can be approximated as independent Gaussian variables because we assumed that the squeezing parameter $\Delta$ is small enough.

\section{Gaussian Shift Error Channel}
\label{sec:Gaussian_shift_error}

As in \cref{eq:approximate_wave_function}, an approximate GKP code state $\ket{\tilde{\psi}}$ can be written as a coherent superposition (up to phase factors):
\begin{align*}
\ket{\tilde{\psi}} = \int \D u \D v \sqrt{P_{u} P_{v}} e^{-i u\hat{p} + i v\hat{q}} \ket{\overline{\psi}}
\end{align*}
where $\ket{\overline{\psi}}$ is the ideal state and $P_{u}, P_{v}$ are probability densities of $u,v$. The error channel transforms an ideal state to an approximate state can be represented by a superoperator $\mathcal{E}$:
\begin{align}
\rho \rightarrow \mathcal{E}(\rho) =\int\D\alpha \D\alpha' \mathcal{P}(\alpha,\alpha') D(\alpha)\rho D^{\dag}(\alpha')
\label{eq:non-twirl}
\end{align}
where $\alpha = u- iv$ and $\mathcal{P}(\alpha,\alpha') = \sqrt{ P_{u}P_{v}P_{u'}P_{v'} }$. Note that $D(\alpha)$ is the displacement operator that:
\begin{align*}
D(\alpha) = \exp(\alpha \hat{a}^{\dag} - \alpha^*\hat{a} ),
\end{align*}
and it's easy to check that $D(\alpha) = \exp(-i u\hat{p} + i v\hat{q})$. Using a generalized Pauli Twirling Approximation as shown in Sec.(\ref{sec:PTA}), a special case of 'Pauli Channel' where only diagonal terms of $P(\alpha,\alpha')$ are left :
\begin{align}
\rho \rightarrow \mathcal{E}(\rho) = \int\mathrm{d}u \mathrm{d}v P_{u} P_{v} D(u,v)\rho D^{\dag}(u,v).
\label{eq:twirl_infinite}
\end{align}

In this thesis, we set the probability densities as Gaussian distributions with variance $\sigma$, i.e. $P_{u} = P_{\sigma}(u)$ and $P_{v} = P_{\sigma}(v)$. And we call this channel as the 'Gaussian Shift Error Channel'. The effect of this channel is that position $q$ and momentum $p$ are displaced by independent Gaussian variables $u,v$ respectively (up to phase factors):
\begin{align*}
q \to q + u  \qquad  p \to p + v
\end{align*}

\subsection{Generalized Pauli Twirling Approximation}
\label{sec:PTA}

First we consider the original Pauli Twirling Approximation of a single qubit \cite{katabarwa2017dynamical}. The time evolution of a density matrix $\rho$ represented by some superoperator $\mathcal{E}'$ is set to be:
\begin{align}
\rho \rightarrow \mathcal{E}'(\rho) =\sum_{n,m=1}^{4} \chi_{m,n} B_m\rho B^{\dag}_n
\label{eq:non_pta}
\end{align}
where $\chi$ is a positive Hermitian matrix and $B_m$ is an element of the Pauli group $\mathcal{P} = \{I,X,Y,Z\}$. Twirling this channel with any possible tensor product of the Pauli group $\mathcal{P}$, written as $\mathbb{B}_k$. We can get a new quantum channel ($K$ is the number of possible tensor products):
\begin{equation}
\begin{split}
\rho \rightarrow \bar{\mathcal{E}}' (\rho) & =\frac{1}{K}\sum_{k=1}^K \mathbb{B}^{\dag}_k \mathcal{E}(\mathbb{B}_k \rho \mathbb{B}^{\dag}_k) \mathbb{B}_k \\
& =\sum_{n=1}^{4} P_{n} B_n\rho B^{\dag}_n
\end{split}
\label{eq:twirl_finite}
\end{equation}
where  the cross-terms in Eq.(\ref{eq:non_pta}) has been eliminated and $P_{n}$ is a probability density. Similarly for the superoperator $\mathcal{E}$ in Eq.(\ref{eq:non-twirl}), we do the Pauli Twirling Approximation with displacement operators $D(\beta)$ with $\beta$ an arbitrary complex number:
\begin{align}
\bar{\mathcal{E}}(\rho) = \frac{1}{\mathcal{N}} \int \mathrm{d}\beta D^{\dag}(\beta)\mathcal{E}\left( D(\beta) \rho D^{\dag}(\beta) \right) D(\beta)
\label{eq:twirl_infinite_unphysical}
\end{align}
where $\mathcal{N}$ is the normalization factor that $\frac{1}{\mathcal{N}}\int \D \beta = 1$. It's easy to check that this gives us exactly the 'Pauli Channel' in Eq.(\ref{eq:twirl_infinite}). However, we cannot really apply the displacement operators with $\beta \to \infty$, we thus need to localize $\beta$ around 0 with a probability distribution:
\begin{align}
\frac{1}{\gamma \pi} \int \mathrm{d}\beta e^{-\frac{\vert \beta \vert^2}{\gamma}} = 1
\label{eq:probability_distribution}
\end{align}
Similarly, we have the new channel after twirling:
\begin{align}
\bar{\mathcal{E}}(\rho) = \frac{1}{\gamma \pi} \int \mathrm{d}\beta e^{-\frac{\vert \beta \vert^2}{\gamma}} D^{\dag}(\beta)\mathcal{E}\left( D(\beta) \rho D^{\dag}(\beta) \right) D(\beta)
\end{align}
With the identity $D(\alpha)D(\beta) = e^{(\alpha\beta^{*}-\beta\alpha^{*})/2}D(\alpha+\beta)$, we have the twirled channel as:
\begin{align}
\bar{\mathcal{E}}(\rho) &= \int \mathrm{d}\alpha \int \mathrm{d}\alpha' P(\alpha, \alpha') D(\alpha) \rho D^{\dag}(\alpha') 
\cdot \underbrace{ \frac{1}{\gamma \pi}  \int \mathrm{d}\beta e^{-\frac{\vert \beta \vert^2}{\gamma}} e^{[(\alpha\beta^{*}-\beta\alpha^{*}) - (\alpha'\beta^{*}-\beta\alpha'^{*})]} }_{ e^{-\gamma \vert \alpha - \alpha' \vert^2} }\\
& = \int \mathrm{d}\alpha D(\alpha) \rho \int \mathrm{d}\alpha' P(\alpha, \alpha') D^{\dag}(\alpha') e^{-\gamma \vert \alpha - \alpha' \vert}, 
\end{align}
assume that $\gamma$ is large enough:
\begin{align}
\gamma \gg \max (\frac{1}{\vert \alpha - \alpha' \vert^2}),
\end{align}
then the terms with $\alpha  \neq \alpha'$ becomes negligible and only the diagonal terms are left (up to normalization):
\begin{align}
\bar{\mathcal{E}}(\rho) \propto \int \mathrm{d}\alpha  P(\alpha) D(\alpha) \rho D^{\dag}(\alpha),
\end{align}
where $P(\alpha) = P(\alpha,\alpha)$ is the diagonal term. This is exactly the "Pauli Channel" in \cref{eq:twirl_infinite}. Note that when $\gamma \to \infty$, the probability distribution in \cref{eq:probability_distribution} becomes a uniform distribution, and the twirling operation is reduced back to the unphysical one in \cref{eq:twirl_infinite_unphysical}.

\subsection{Comparing Two Kinds of Shift Errors}

Here we compare the error from the Gaussian shift error channel with the internal error due to finite squeezing.
First we prepare a qubit in a finitely squeezed state $\ket{\tilde{\psi}}$ and assume the errors in both quadrature satisfy a Gaussian distribution with variance $\sigma$:
\begin{align}
\ket{\tilde{\psi}} = \int \D u \D v \sqrt{P_{\sigma}(u)P_{\sigma} (v)}\cdot e^{-iu\hat{p}}e^{iv\hat{q}}\ket{\overline{\psi}},
\label{eq:coherent_error}
\end{align}
where $\ket{\tilde{\psi}}$ is a superposition of ideal states subjected to shift errors. On the other hand, we prepare another qubit in the ideal state $\ket{\overline{\psi}}$ and then let it go through the Gaussian shift error channel as in \cref{eq:twirl_infinite} with the same variance $\sigma$. The state of this qubit is a mixed state, described by a density matrix $\tilde{\rho}$:
\begin{align}
\tilde{\rho} = \int \D u \D v P_{\sigma}(u)P_{\sigma} (v)\cdot e^{-i u\hat{p}}e^{+i v\hat{q}}\ket{\overline{\psi}} \bra{\overline{\psi}} e^{+i u\hat{p}}e^{-i v\hat{q}}.
\label{eq:stochastic_error}
\end{align}
The above two states are different, i.e. one is in pure state and the other is in mixed state. But if the shift errors are measured perfectly, we actually cannot tell the difference between them. After perfectly measuring the values of $u,v$, we will find both qubits are in state $\ket{\psi_{out}}$ subjected to some shift errors $u_0,v_0$:
\begin{align}
\ket{\psi_{out}} = e^{-i u_0 \hat{p}}e^{i v_0 \hat{q}}\ket{\psi},
\label{eq:post-measure_state}
\end{align}
with the same probability density $P(u_0,v_0) = P_{\sigma}(u_0)P_{\sigma}(v_0)$. Obviously, we cannot tell the difference between these two qubits after measuring shift errors, if we only have the information of the measurement outcomes $u_0, v_0$.

However, the above two qubits would be in different states after noisy measurements. \cref{eq:coherent_error} is an example of coherent shift error, and \cref{eq:stochastic_error} is an example of stochastic shift error. Following in this thesis we assume ideal qubits subjected to stochastic shift errors, and we expect the coherent shift error would produce similar effects, which needs to be analyzed further.

\chapter{Quantum Computation with GKP Code States}
\label{sec:QC-GKP}

This chapter is about some basic applications of GKP code states in quantum computation. Section \ref{sec:clifford gate} is about the effects of Clifford gates acting on GKP code states, which correspond to symplectic operations, i.e. linear transformations of the quadratures of an oscillator, preserving the canonical commutation relations. Also in this section we examine the error propagations through these gates. \cref{sec:universal-qc} is about realizing universal quantum computation utilizing photon counting, where Clifford gates are simple to implement because they only involve linear optical elements\cite{gottesman2001encoding}. In Section \ref{sec:steane_qec}, a quantum error correction protocol named Steane error correction is introduced, which enables us to efficiently correct the shift errors preventing them from accumulating. The last Section \ref{sec:MBQ_gkp} is about GKP code states in continuous variable cluster states and measurement-based quantum computation (MBQ). 
\section{Clifford Gates of GKP code states}
\label{sec:clifford gate}

The group of Clifford gates (CNOT, phase and Hadamard gates) is an important subgroup of gates. In general, the Clifford group of a system of N qubits is the group of unitary transformations that, acting by conjugation, takes tensor products of Pauli operators to tensor products of Pauli operators\cite{gottesman2001encoding}:
\begin{align*}
\text{CNOT}: \quad &X_1\to X_1 X_2 && Z_1\to Z_1\\
&X_2\to X_2 && Z_2 \to Z_1 Z_2\\
\text{Hadamard}\quad H: \quad & X \to Z && Z\to X\\
\text{Phase}\quad  S: \quad & X \to i X Z && Z\to Z
\end{align*}
where the subscripts 1,2 correspond to the control and target qubit of the CNOT gate respectively.
From the Gottesman-Knill theorem\cite{nielsen2000quantum}, we know that quantum computation involving only the Clifford gates and Pauli operators can be efficiently simulated by classical computation, so the Clifford group is not enough for universal quantum computation. This will be examined later in Section \ref{sec:universal-qc}.

\subsection{Action on the GKP Code States}

Now we first take a look at Clifford Gates acting on GKP code states. In general for a system consisted of N oscillators, the tensor products of displacement operators can be expressed in terms of the canonical quadratures $\hat{q}_i$ and $\hat{p}_i$ as:
\begin{align*}
U_{\alpha\beta} = \exp \left[\mathrm{i}\sqrt{2\pi}\left( \sum\limits_{i=1}^{N} \alpha_i \hat{p}_i + \beta_i \hat{q}_i  \right) \right],
\end{align*}
where the $\alpha_i$ and $\beta_i$ are real numbers. A special case for a single GKP code encoded into an oscillator, the logical Pauli operators are :
\begin{align*}
\bar{X} = e^{i \sqrt{\pi}\hat{p} },\qquad \bar{Z} = e^{i \sqrt{\pi} \hat{q} }.
\end{align*}
For GKP encoded qubits, all Clifford gates must be symplectic operations which are linear transformations of the $p$'s and $q$'s that preserve the canonical commutation relations $[\hat{p},\hat{q}]$, acting by conjugation. And it's easy to determine the symplectic operations corresponding to the Clifford gates as shown below\cite{gottesman2001encoding}:
\begin{align*}
\text{CNOT}: \quad &\hat{q}_1\to \hat{q}_1 && \hat{p}_1\to \hat{p}_1+\hat{p}_2\\
&\hat{q}_2\to \hat{q}_2-\hat{q}_1 && \hat{p}_2 \to \hat{p}_2\\
\text{Hadamard} \quad H: \quad &\hat{q} \to \hat{p} && \hat{p} \to -\hat{q}\\
\text{Phase}\quad S: \quad &\hat{q} \to \hat{q} && \hat{p} \to \hat{p} -\hat{q}
\end{align*}
For the GKP code states, the group generated by Clifford gates is a subgroup that is "easy" to implement, since these gates only requires linear optical elements (phase shifts and beam splitters) along with elements that can "squeeze" an oscillator \cite{gottesman2001encoding} \cite{glancy2006error}.

\subsection{Error Propagation}
\label{sec:error propagation}

Considering that Clifford gates on GKP code states are linear transformations of the $\hat{q}$ quadrature and the $\hat{p}$ quadrature, these gates can move shift errors from one qubit to another or from one quadrature to another. Here we examine the error propagations of Clifford gates and take the CNOT gate as an example.
\begin{figure}
\centering
\begin{minipage}{\textwidth}
\[
\Qcircuit @C=2.0em @R=1.4em {
\lstick{e^{-\mathrm{i} u_1 \hat{p}_1} e^{i v_1 \hat{q}_1} \ket{\overline{\psi} } }  & \ctrl{1}  &\rstick{e^{-i u_1 \hat{p}_1} e^{\mathrm{i} (v_1-v_2) \hat{q}_1} \ket{\overline{\psi}} }\qw\\
\lstick{e^{-i u_2 \hat{p}_2}e^{i v_2 \hat{q}_2} \ket{\overline{+}} }  & \targ             &\rstick{e^{-i (u_2+u_1) \hat{p}_2} e^{i v_2 \hat{q}_2}\ket{\overline{+}} } \qw
}\]
\end{minipage}
\caption{Circuit of the CNOT gate, where $\ket{ \overline{\psi} }$ represents an arbitrary ideal GKP code state. The logical CNOT gate moves the shift error $u_1$ of $\ket{ \overline{\psi} }$ to $\ket{ \overline{+} }$, and moves the shift error $v_2$ in the opposite direction.}
\label{fig:CNOT_Error}
\end{figure}

Assume the control qubit is in ideal state $\ket{\overline{\psi}_1}$ and target qubit is in ideal state $\ket{\overline{\psi}_2}$, subjected to shift errors $u_1$, $u_2$ in the $\hat{q}$ quadrature and $v_1$,$v_2$ in the $\hat{p}$ quadrature respectively (up to phase factors):
\begin{align*}
\ket{\psi_{in}} \propto e^{\mathrm{i} (u_1 \hat{p}_1+u_2 \hat{p}_2)} e^{\mathrm{i}( v_1 \hat{q}_1+v_2 \hat{q}_2)}\ket{\bar{\psi}_1,\bar{\psi}_2 }.
\end{align*}
Recall that CNOT ($C_x$) transforms $\hat{p}_1 \to \hat{p}_1+\hat{p}_2$ and $\hat{q}_2 \to \hat{q}_2 - \hat{q}_1$, then the effect of this circuit in Fig(\ref{fig:CNOT_Error}) is (up to some phase factors):
\begin{align*}
C_x: \quad \ket{\psi_{in}} & \to C_x e^{i (u_1 \hat{p}_1+u_2 \hat{p}_2)} e^{i( v_1 \hat{q}_1+v_2 \hat{q}_2)} C_x^{\dag} \cdot C_x \ket{\bar{\psi}_1,\bar{\psi}_2 } \\
&\propto e^{i [u_1 (\hat{p}_1+\hat{p}_2) +u_2 \hat{p}_2])} e^{i[v_1 \hat{q}_1+v_2 (\hat{q}_2-\hat{q}_1)]}\cdot C_x  \ket{\bar{\psi}_1,\bar{\psi}_2 } \\
&\propto e^{\mathrm{i} [u_1 \hat{p}_1+( v_1-v_2) \hat{q}_1]} e^{\mathrm{i}[(u_2+u_1) \hat{p}_2 + v_2 \hat{q}_2]} \cdot C_x  \ket{\bar{\psi}_1} \ket{\bar{\psi}_2 }.
\end{align*}
It's clear that the CNOT gate moves control qubit's shift error $u_1$ in the $\hat{q}$ quadrature to the target qubit. It also propagates the shift error $v_2$ in the $\hat{p}$ quadrature from target to control qubit. 

For simplicity, we demonstrate the effect of the error propagation with the target qubit in ideal state $\ket{\overline{+}}$ subjected to shift error $u_2,v_2$， as shown in \cref{fig:CNOT_Error}, .

\section{Universal Quantum Computation}
\label{sec:universal-qc}

The Clifford gates discussed above in \cref{sec:clifford gate} are not enough for universal quantum computation \cite{nielsen2000quantum}. To complete a set of gates to achieve universal quantum computation, an additional gate is needed, for example the $\pi/8$ gate. Such a gate requires non-symplectic operations. Here we first use Hadamard eigenstates to realize the $\pi/8$ gate, and then use photon counting to prepare such an eigenstate proposed by Gottesman \textit{et al.} \cite{gottesman2001encoding}.

\subsection{The $\pi/8$ Gate using Hadamard Eigenstate}
\begin{figure}
\centering
\begin{minipage}{\textwidth}
\[
\Qcircuit @C=1.2em @R=0.8em {
\lstick{\ket{\phi } } &\qw & \ctrl{1}  &\gate{S} & \rstick{ T \ket{\phi } } \qw\\
\lstick{T\ket{+} } &\qw  & \targ             &\measureD{Measure} \cwx[-1]
}\]
\end{minipage}
\caption{Circuit for gate teleportation. The data qubit is in an arbitrary state $\ket{\phi }$. The ancilla qubit is in state $\ket{\Phi_{\pi/8}}= T\ket{+}$. Measure the ancilla in the basis $ {\ket{0},\ket{1}} $, and apply a conditional phase gate S on the data qubit. The output qubit is in state $ T \ket{\phi }$ up to some phase factors.}
\label{fig:gate_teleport}
\end{figure}
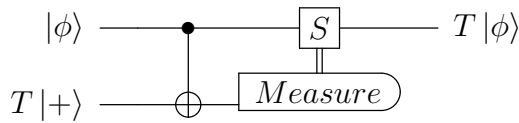

Assume that we already prepared a qubit in the Hadamard eigenstate corresponding to eigenvalue $+1$:
\begin{align*}
\ket{\Psi_{\mathbb{H}=1}} = \cos(\pi/8) \ket{0} + \sin(\pi/8) \ket{1}.
\end{align*}
Apply the symplectic transformation:
\begin{align*}
H \cdot S^{-1} = \frac{1}{\sqrt{2}}
\left( 
\begin{array}{cc}
1 & 1 \\
1 & -1 
\end{array} 
\right )
\cdot
\left( 
\begin{array}{cc}
1 & 0 \\
0 & -i 
\end{array} 
\right )
= \frac{1}{\sqrt{2}}
\left( 
\begin{array}{cc}
1 & -i\\
1 & i 
\end{array} 
\right ).
\end{align*}
We can thus obtain a so called $\pi/8$ phase state:
\begin{align*}
\ket{\Psi_{\pi/8} } = \frac{1}{\sqrt{2}} ( e^{-\mathrm{i}\pi/8} \ket{0} + e^{\mathrm{i}\pi/8} \ket{1} ).
\end{align*}
This phase state enables us to perform a non-symplectic  gate:
\begin{equation*}
T =
{
\left( \begin{array}{cc}
e^{-\mathrm{i}\pi/8} & 0 \\
0 & e^{\mathrm{i}\pi/8} 
\end{array} 
\right )
}.
\end{equation*}
$T$ is the $\pi/8$ phase gate, and the state $\ket{\Psi_{\pi/8} }$ can be written as applying $T$ gate on state $\ket{+}$:
\begin{align*}
\ket{\Psi_{\pi/8} } = T \ket{+}
\end{align*}
Then we can utilize the gate teleportation circuit to teleport $T$ gate from the $\pi/8$ phase state to an arbitrary qubit state $\ket{\phi }$ \cite{gottesman2001encoding} \cite{nielsen2000quantum}, which completes the set of gates that realizes universal quantum computation. The circuit for gate teleportation is shown in \cref{fig:gate_teleport}.

\subsection{Photon Number Modulo Four}
\label{sec:photon_counting}

In order to see the relation between photon counting and the Hadamard eigenstate, we first write the Hamiltonian $\mathcal{H}$ of the oscillator as:
\begin{align}
\mathcal{H} = \hbar \omega (\hat{a}^{\dag} \hat{a}+\frac{1}{2}),
\end{align}
where $\hbar$ is the reduced Plank constant and $\omega$ a real constant. The time evolution operator with time $t$ is thus (up to phase factors):
\begin{align}
U(t) = e^{-\frac{i}{\hbar} \mathcal{H} t} = e^{-i \omega t \hat{a}^{\dag} \hat{a}} .
\end{align}
The effect of this time evolution operator on $\hat{q}$ and $\hat{p}$ can be written as:
\begin{equation}
\begin{split}
\hat{q} &\to U^{\dag}(t) \hat{q } U(t)  \propto   \hat{a}e^{-i \omega t} + \hat{a}^{\dag} e^{i \omega t},\\
\hat{p} &\to U^{\dag}(t) \hat{p } U(t) \propto  \hat{a}e^{-i \omega t} - \hat{a}^{\dag} e^{i \omega t}.
\end{split}
\label{eq:qp_evolution}
\end{equation}
It's clear, when the evolution time $t = \frac{\pi}{2\omega}$, the effect of \cref{eq:qp_evolution} is just the Hadamard gate that $ H^{\dag} \hat{q} H \to \hat{p}$ and $ H^{\dag} \hat{p} H \to -\hat{q}$, see \cref{sec:clifford gate}. Thus the Hadamard gate represents a quarter cycle of the time evolution \cite{gottesman2001encoding}:
\begin{align*}
H: \quad \exp \left( i \frac{\pi}{2} \hat{a}^{\dag}\hat{a} \right),
\end{align*}
where the phase is simply the photon number modulo four, thus the $+1$ eigenstate of Hadamard gate is a state with photon number equal to $0 \mod 4$.

In the quadrature plane of the GKP code, all code words are invariant under a $\pi$ rotation ( $\hat{q} \to -\hat{q}, \hat{p} \to -\hat{p}$), which is easy to check from the definition of the code words in \cref{eq:01_q_quadrature} and \cref{eq:+-_q_quadrature}. Note that the $\pi$ rotation is exactly a time evolution operator with $t = \frac{\pi}{\omega}$, check it in \cref{eq:qp_evolution}. Thus an arbitrary GKP code state with arbitrary photon number $n$ should be an eigenstate of the operator $U(t = \pi/ \omega)$:
\begin{align*}
U(t = \pi/ \omega) \ket{n} = e^{i n \pi} \ket{n} = \ket{n}.
\end{align*}
It's clear that the photon number $n$ can only be even to satisfy the equation above. This restriction of photon number gives us some fault-tolerance measuring it.

\subsection{Preparing a Hadamard Eigenstate}

In the non-demolition protocol of photon counting proposed by Gottesman \textit{et al.}\cite{gottesman2001encoding}, the photon number of an encoded state is measured repeatedly. First, we couple the oscillator to an atom with a chosen perturbation:
\begin{align*}
\mathcal{H}' = \hbar \omega' \hat{a}^{\dag} \hat{a} \sigma_z,
\end{align*}
where $\sigma_z \ket{0}_a = -\ket{0}_a$ with $\ket{0}_a$ the atomic ground state, $\sigma_z \ket{1}_a = \ket{1}_a$ with $\ket{1}_a$ the excited state $\ket{1}_a$. By turning on the coupling for a time $t = \frac{\pi}{4\omega'}$, the executed unitary operator is a time evolution operator with $t = \pi/(4\omega')$:
\begin{align*}
U(t = \frac{\pi}{4\omega'}) = \exp[-i (\pi/4)\hat{a}^{\dag} \hat{a} \sigma_z ].
\end{align*}
Then the coupled system $\frac{1}{\sqrt{2}} (\ket{0}_a+\ket{1}_a) \ket{n}$ evolves as:
\begin{align*}
\frac{1}{\sqrt{2}}(\ket{0}_a+\ket{1}_a) \ket{n} &\to
U(t = \frac{\pi}{4\omega'})\cdot \frac{1}{\sqrt{2}} (\ket{0}_a+\ket{1}_a) \ket{n} \\
& = \frac{1}{\sqrt{2}}(e^{i n \pi/4}\ket{0}_a+ e^{-i n \pi/4}\ket{1}_a) \ket{n}\\
&=  \frac{1}{\sqrt{2}} e^{i n \pi/4}\cdot (\ket{0}_a+ e^{-i n \pi/2}\ket{1}_a) \cdot \ket{n}.
\end{align*}

As discussed in \cref{sec:photon_counting}, the photon number can only be $0 \text{or} 2 \mod 4$. If $n = 0 \mod 4$, the atomic state will be left in state $\frac{1}{\sqrt{2}} (\ket{0}_a+\ket{1}_a)$, otherwise in state $\frac{1}{\sqrt{2}} (\ket{0}_a-\ket{1}_a)$. Thus by measuring the atomic state in the basis $(\ket{0}_a\pm \ket{1}_a)/\sqrt{2}$, we read out the value of the photon number modulo four of the GKP code state, also we know that we have prepared the qubit in a Hadamard eigenstate with a known eigenvalue.

Note that this measurement is non-demolition, which can be repeated to improve reliability. Repeating measurements can increase the fidelity of a Hadamard eigenstate nicely.

\section{Steane Error Correction Scheme}
\label{sec:steane_qec}

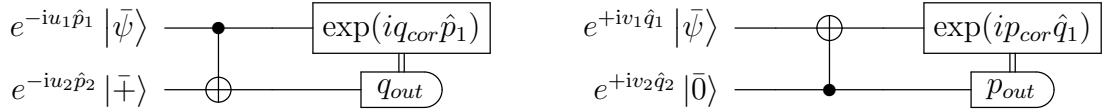
\begin{figure}
\centering
\begin{minipage}{\textwidth}
\[
\Qcircuit @C=1.3em @R=0.6em {
\lstick{ e^{-\mathrm{i} u_1 \hat{p}_1} \ket{ \bar{\psi} } }  & \ctrl{1} & \qw &\gate{\exp(i q_{cor} \hat{p}_1)} 
&&&&&&
\lstick{e^{+\mathrm{i} v_1 \hat{q}_1} \ket{ \bar{\psi} } }&\qw &\targ  & \qw &\gate{\exp(i p_{cor} \hat{q}_1)}
\\
\lstick{e^{-\mathrm{i} u_2 \hat{p}_2} \ket{ \bar{+} } }  & \targ & \qw  &\measureD{q_{out}} \cwx[-1] 
&&&&&&
\lstick{e^{+\mathrm{i} v_2 \hat{q}_2} \ket{\bar{0} } } &\qw  &\ctrl{-1} & \qw  &\measureD{p_{out}} \cwx[-1]
}\]
\end{minipage}
\caption{Circuits of the Steane error correction scheme. The left circuit corrects shift errors in the $\hat{q}$ quadrature and right one in the $\hat{p}$ quadrature. The logical CNOT moves the shift errors as discussed in \cref{sec:error propagation}. We have incoming shift errors $u_1,v_1 \in [-\sqrt{\pi}/2,\sqrt{\pi}/2]$. And ancilla qubits $\ket{\bar{+}},\ket{\bar{0}}$ have shift error $u_2,v_2 \in [-\sqrt{\pi}/2,\sqrt{\pi}/2]$. After the CNOT gate, we measure the ancillas in $q,p$ quadratures respectively and get the homodyne measurement outcomes $q_{out} = u_1 + u_2 + n_1 \sqrt{\pi}$ and $ p_{out} = -v_1 + v_2 + n_2 \sqrt{\pi}$ with $n_1, n_2$ arbitrary integers. Then we calculate $q_{cor} = q_{out} \mod \sqrt{\pi} \in [-\sqrt{\pi}/2,\sqrt{\pi}/2]$ and apply the correction operators, which is the same for $p_{cor}$.
}
\label{fig:steane_circuit}
\end{figure}

Steane error correction corrects small shifts in position or momentum quadratures fault-tolerantly\cite{gottesman2001encoding} \cite{steane1996error} \cite{terhal2016encoding}. Its circuit involves CNOT gates (or beam splitters\cite{terhal2016encoding} \cite{glancy2006error}) and homodyne measurements. We first consider Steane error correction in the $\hat{q}$ quadrature, the left one in \cref{fig:steane_circuit}. The analysis of the $\hat{p}$ quadrature is completely the same, since we assume two quadratures are symmetric and shift errors in them are independent.

We assume that the input state has shift error $u_1$ and the ancilla is in state $\ket{\bar{+}}$ subjected to shift error $u_2$. Up to normalization, the initial state of the combined system of an input qubit and an ancilla qubit is:
\begin{align}
&e^{-iu_1\hat{p}_1}e^{-iu_2\hat{p}_2} \ket{\bar{\psi}}\ket{\bar{+}}.
 \label{eq:qec_steane_initial}
\end{align}
The circuit to correct shifts in the $\hat{q}$ quadrature is shown in \cref{fig:steane_circuit}. We already discussed how the CNOT gate moves  the shift errors in \cref{sec:error propagation}, it's easy to check that the CNOT gate maps the combined system to a new state:
\begin{align}
 e^{-iu_1\hat{p_1}}e^{-iu_2\hat{p_2}}\ket{ \overline{\psi} }_1\ket{ \overline{+} }_2 \to e^{-iu_1\hat{p_1}}e^{-i(u_1+u_2)\hat{p_2}}\ket{ \overline{\psi} }_1\ket{ \overline{+} }_2,
\label{eq:steane_q_state_evolution}
\end{align}
where the subscript 1,2 represent control and target qubit of the CNOT gate respectively. With the definition of ideal GKP code states as in \cref{eq:01_q_quadrature}, we can write the output state as:
\begin{equation}
\begin{split}
=\sum_{s,s'\in \mathbb{Z}}\  &\alpha \ket{q_1 = 2s\sqrt{\pi}+u} \ket{q_2=(2s+s')\sqrt{\pi}+u} \\
 &+ \beta \ket{q_1=(2s+1)\sqrt{\pi}+u} \ket{q_2=(2s+1+s')\sqrt{\pi}+u}\\
= \sum_{n_0,n_1 \in \mathbb{Z}} &\Big( \alpha \ket{q_1 = 2n_0\sqrt{\pi}+u_1}+\beta \ket{q_1=(2n_0+1)\sqrt{\pi}+u_1}\Big) \ket{q_2=n_1\sqrt{\pi}+u_1+u_2},
\end{split}
\label{eq:output_steane_q}
\end{equation}
where we write $\ket{\bar{\psi}} = \alpha \ket{\bar{0}} + \beta \ket{\bar{1}}$ with $\vert \alpha \vert^2 + \vert \beta \vert^2 = 1$. 

The measurement of the ancilla can only produce $q_{out} = u_1 + u_2 + n_1 \sqrt{\pi}$, $n_1\in \mathbb{Z}$. Such a measurement reveals no information about $\alpha$ or $\beta$, thus does not destroy the input state. Then we calculate $q_{cor} = q_{out} \mod \sqrt{\pi} \in  [-\sqrt{\pi}/2,\sqrt{\pi}/2]$ and apply the correction operator $\exp(i q_{cor} \hat{p})$, the input qubit $\ket{\bar{\psi}}$ is now in state (neglecting some phase factors):
\begin{align*}
&\sum_{s \in \mathbb{Z}} \Big( \alpha \ket{q_1 = 2s\sqrt{\pi} - u_2 - n_1\sqrt{\pi}}+\beta \ket{q_1=(2s+1)\sqrt{\pi}-u_2 -n_1\sqrt{\pi}}\Big)\\
&=  e^{\mathrm{i}n_1 \sqrt{\pi}\hat{p}} e^{\mathrm{i}u_2 \hat{p}}\ket{\bar{\psi}},
\end{align*}
that is, according to the measurement outcome the code states are shifted to the nearest integer multiple of $\sqrt{\pi}$ plus a small shift error $u_2$ from the ancilla. Note that $e^{i n_1 \sqrt{\pi} }$ is a logical $\bar{X}$ operator when $n_1$ is odd and otherwise an identity operator, see \cref{sec:encoding_oscillator}. Also the analysis in the $\hat{p}$ quadrature is the same, where the CNOT gate is inverted, and an ancilla initially prepared in $\ket{\bar{0}}$ is measured in the $\hat{p}$ quadrature at the end, see \cref{fig:steane_circuit}.

For Steane error correction in the $\hat{q}$ quadrature, the whole effect on the data qubit's shift errors is (up to possible logical shifts in both quadratures):
\begin{equation}
 u_1 \to - u_2, \quad v_1 \to v_1-v_2,
 \label{eq:steane_q_error}
\end{equation}
with success condition $\vert u_1+u_2\vert < \frac{\sqrt{\pi}}{2}$. In the $\hat{p}$ quadrature, the effect is :
\begin{equation}
u_1 \to u_1 + u_2, \quad  v_1 \to v_2,
\label{eq:steane_p_error}
\end{equation}
with success condition $\vert v_2-v_1\vert <\frac{\sqrt{\pi}}{2}$. Where $u,v$ are shift errors in the $\hat{q}$ and the $\hat{p}$ quadratures respectively. Subscript 1,2 represents data qubit and ancilla qubit. Note that a minus sign doesn't matter, because for an arbitrary Gaussian variable $u$ with mean value zero, $u$ and $-u$ have completely the same probability distribution.

\subsection{Steane Error Correction is Fault-Tolerant}
\label{sec:fault-tolerant_Steane}
For repeated Steane error corrections are applied, Glancy and Knill\cite{glancy2006error} found an error threshold equal to $\sqrt{\pi}/6$, under which there's always no logical error. We first do Steane error correction in the $\hat{q}$ quadrature as the left circuit in \cref{fig:steane_circuit}, followed by Steane error correction in the $\hat{p}$ quadrature as the right one in \cref{fig:steane_circuit}. We call the procedure described above as a round of correction and repeat it.

It is assumed that data qubit contains shift error $u_1$ and $v_1$ and an ancilla subjected to shift error $u_2,v_2$. Since the effect of Steane error correction is shown in \cref{eq:steane_q_error} and \cref{eq:steane_p_error}, thus after Steane error correction in the $\hat{q}$ quadrature, we have:
\begin{align*}
u_1,v_1 \to  -u_2, v_1-v_2,
\end{align*}
with success condition $\vert u_1 + u_2 \vert < \sqrt{\pi}/2$. In the following correction in the $\hat{p}$ quadrature with an ancilla subjected to shift error $u_3,v_3$:
\begin{align*}
-u_2, v_1-v_2 \to -u_2+u_3, v_3,
\end{align*}
with success condition $\vert -v_1 +v_2+v_3\vert < \sqrt{\pi}/2$. Thus the first round of error correction is complete, the qubit will be left with shift error $-u_2+u_3$ and $v_3$ in the $\hat{q}$ and the $\hat{p}$ quadratures respectively. In the second round, the correction in the $\hat{q}$ quadrature, with an ancilla subjected to shift error $u_4,v_4$, transforms the errors as:
\begin{align*}
-u_2+u_3, v_3 \to -u_4, v_3-v_4,
\end{align*}
the success condition is $\vert -u_2 + u_3 + u_4\vert <\sqrt{\pi}/2$. In the second correction in the $\hat{p}$ quadrature with an ancilla subjected to shift error $u_5,v_5$, we have:
\begin{align*}
-u_4,v_3-v_4 \to -u_4+u_5, v_5,
\end{align*}
with success condition $\vert -v_3+v_4+v_5\vert <\sqrt{\pi}/2$. If we repeat the correction procedure described above, each time the qubit has error inherited from one ancilla in the quadrature that was just corrected, and errors from two ancillas in the other quadrature that will be corrected next. 

Thus the magnitude of the three shift errors should always be smaller thatn $\sqrt{\pi}/2$ in order to ensure that correction always succeeds. For a single error from data or ancilla qubit, the error threshold threshold of it is exactly $\sqrt{\pi}/6$. Under this threshold there will never be an undetectable logical error, which shows that Steane error correction is fault-tolerant, as we mentioned in \cref{sec:intrinsic_fault_tolerant}

\subsection{Steane Error Correction with 1-Bit Teleportation}

\begin{figure}
\centering
\begin{minipage}{\textwidth}
\[
\Qcircuit @C=1.3em @R=0.6em {
\lstick{ e^{-\mathrm{i}u_1 \hat{p}_1} \ket{ \bar{\psi} } }  & \targ &\measureD{q_{out}} 
&&&&&&&&
\lstick{ e^{-\mathrm{i}v_1 \hat{q}_1}\ket{ \bar{\psi}} } &\ctrl{1}  &\measureD{p_{out}} 
\\
\lstick{ e^{-\mathrm{i}u_2 \hat{p}_2}\ket{ \bar{+} } }  & \ctrl{-1} & \qw  
&&&&&&&&
\lstick{ e^{-\mathrm{i}v_2 \hat{q}_2}\ket{ \bar{0} } }  &\targ & \qw   
}\]
\end{minipage}
\caption{Circuits for Steane error corrections via 1-bit teleportation. Left circuit corrects errors in the $\hat{q}$ quadrature. After the CNOT gate we measure the data qubit in the $\hat{q}$ quadrature with outcome $q_{out} = u_1 + u_2 + n \sqrt{\pi}, n\in \mathbb{Z}$. Based on $q_{out}$, we need to decide the state of data qubit after measurement, i.e. whether $q_{out}$ is closer to an even multiple of $\sqrt{\pi}$ or closer to an odd multiple. If it's in $\ket{0}$, the ancilla will be left in state $\ket{\bar{\psi}}$ or with shift error $u_2$ and $v_1 + v_2$, otherwise there would be a logical error. The analysis for the left circuit  which corrects errors in the $\hat{p}$ quadrature is the same. 
}
\label{fig:steane_teleport}
\end{figure}
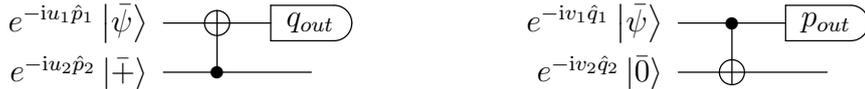

Now we consider a slightly different error correction scheme, which utilizes 1-bit teleportation\cite{zhou2000methodology} as shown in \cref{fig:steane_teleport}). The circuits are quite similar to the circuit of Steane Error correction, but the CNOT gate is inverted and the homodyne measurement is on the input qubit instead.

We first consider correction in the $\hat{q}$ quadrature, i.e the left circuit in Fig.(\ref{fig:steane_teleport}). The input state of the combined system is the same as in Eq.(\ref{eq:qec_steane_initial}), here we write $\ket{\bar{+}} = \frac{1}{\sqrt{2}} (\ket{\bar{0}}+\ket{\bar{1}})$ (neglecting the normalization and phase factors):
\begin{align*}
  & \sum_{s,s' \in \mathbb{Z} } \Big( \alpha \ket{q_1 = 2s\sqrt{\pi}+u_1+u_2} + \beta \ket{q_1=(2s+1)\sqrt{\pi}+u_1+u_2} \Big) \ket{q_2=2s'\sqrt{\pi}+u_2} \\
+ & \sum_{s,s' \in \mathbb{Z} } \Big( \alpha \ket{q_1 = (2s+1)\sqrt{\pi}+u_1+u_2} + \beta \ket{q_1= 2s\sqrt{\pi}+u_1+u_2} \Big) \ket{q_2=(2s'+1)\sqrt{\pi}+u_2}
\end{align*}
we rewrite the state as:
\begin{align*}
\sum_{s,s' \in \mathbb{Z} } \Big( & \ket{q_1 = 2s\sqrt{\pi}+u_1+u_2} \left( \alpha \ket{q_2=2s'\sqrt{\pi}+u_2} + \beta \ket{q_2=(2s'+1)\sqrt{\pi}+u_2}\right) \\
+ &\ket{q_1=(2s+1)\sqrt{\pi}+u_1+u_2} \left( \beta \ket{q_2=2s'\sqrt{\pi}+u_2} +\alpha \ket{q_2=(2s'+1)\sqrt{\pi}+u_2}\right) \Big)
\end{align*}
which can still be simplified to be written as:
\begin{align*}
\sum_{s \in \mathbb{Z} } \Big( \ket{q_1 = 2s\sqrt{\pi}+u_1 + u_2} e^{-\mathrm{i}u_2\hat{p}_2}\ket{\bar{\psi}} + \ket{q_1=(2s+1)\sqrt{\pi}+u_1+u_2} e^{-\mathrm{i}u_2\hat{p}_2} \bar{X}\ket{\bar{\psi}} \Big).
\end{align*}

It's obvious that the measurement of the input qubit in the $\hat{q}$ quadrature produces the measurement outcome $q_{out} = n \sqrt{\pi}+u_1 + u_2$, $n \in \mathbb{Z}$. The next step is to decide the parity of $n$. If $n$ is even, then the ancilla is left in the state $\ket{\bar{\psi}}$ with shift error $u_2$. Otherwise the output ancilla contains a logical $\bar{X}$ error with $n$ odd. Once we make the correct decision, the logical error won't be a serious problem, we can just keep track of and correct it anytime we want\cite{terhal2015quantum}. But with a wrong decision about the parity, we will keep an incorrect record of the Pauli frame.

It's easy to see that the success condition is the same as that of Steane Error Correction: $ \vert u_1 + u_2 \vert \leq \sqrt{\pi}/2$. This scheme has the same effect as the original Steane Error Correction: the input shift error is replaced by the ancillas' shift errors, and contain logical errors with conditional error rates depending on the measurement outcomes.

\section{Measurement-Based Quantum Computation with GKP Code}
\label{sec:MBQ_gkp}

The GKP code fits nicely with fault-tolerant measurement-based quantum computation. By concatenating and using ancilla-based error correction, fault-tolerant measurement-based quantum computation of theoretically indefinite length is possible with finitely squeezed cluster states\cite{Menicucci_Fault_2014} \cite{Menicucci_Supplement_Fault_2014}.

\begin{figure}
\centering
\includegraphics[width=0.6\textwidth]{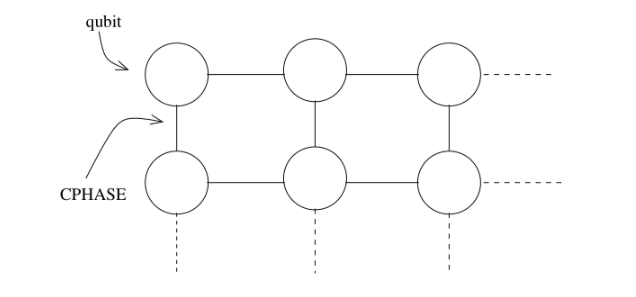}
\caption{A two-dimensional cluster state. Blank nodes means qubits in state $\ket{+}$, the lines connecting qubits represent CPHASE gates.}
\label{fig:cluster-state}
\end{figure}

A typical two-dimensional cluster state can be shown as in \cref{fig:cluster-state}. In order to prepare a cluster state, we initialize each qubit into state $\ket{+}$ ( eigenstate of $\hat{p}$ for GKP code states) and perform CPHASE gates between every pair of neighboring qubits. Since CPHASE gates on different pairs of qubits commute to each other, thus we can apply these gates in parallel. Note that a cluster state is a highly entangled state, which is the source of its ability to perform quantum computation.
It's easy to check that the cluster state is a stabilizer code stabilized by\cite{raussendorf2003measurement}:
\begin{align*}
K^{(a)} = \sigma_x^{(a)} \bigotimes\limits_{b \in \text{nghb}(a)} \sigma_z^{(b)},
\end{align*}
where nghb(a) are all the qubits connected with qubit $a$ by CPHASE gates. A cluster state is completely specified by the eigenvalues of $K^{(a)}$ acting on the cluster state. In order to do quantum computation with a prepared cluster state, we measure qubits in chosen basis to realize desired quantum gates\cite{raussendorf2003measurement}.

Here we concatenate a cluster state with the GKP code, replacing each node by a GKP-encoded qubit prepared in $\hat{p}$ squeezed state, i.e eigenstate of $\hat{p}$. For such a continuous variable cluster state, any single-mode Gaussian unitary can be implemented on a linear CV cluster state consisting of four nodes, measuring node j in the quadrature $\{\hat{p} + m_j \hat{q}\}_{j=1}^4$\cite{Menicucci_Fault_2014}. A quantum gate $\mathbf{G}$ corresponds to a specific measurement vectors defined as $\mathbf{m}^{(\mathbf{G})}                                                                                                                                                                                                                                                                                                                                                                                                                                                                                                                                                                                                                                                                                                                                                                                                                                                                                                                                                                                                                                                                                                                                                                                                                                                                                                                                                                                                                                                                                                                                                                                                                                                                                                                                                                                                                                                                                                                                                                                                                                                                                                                                                                                                                                                                                                                                                                                                                                                                                                                                                                                                                                                                                                                                                                                                                                                                                                                                                                                                                                                                                                                                                                                                                                                                                                                                                                                                                                                                                                                                                                                                                                                                                                                                                                                                                                                                                                                                                                                                                                                                                                                                                                                                                                                                                                                                                                                                                                                                                                                                                                                                                                                                                                                                                                                                                                                                                                                                                                                                                                                                                                                                                                                                                                                                                                                                                                                                                                                                                                                                                                                                                                                                                                                                                                                                                                                                      = (m_1,m_2,m_3,m_4)$. Note that the CNOT gate requires a two dimensional cluster state, see details in the paper of Menicucci\cite{Menicucci_Fault_2014}.

\subsection{Steane Error Correction with Cluster States}
Here we introduce  how to do Steane Error corrections with cluster state concatenated with the GKP code\cite{Menicucci_Supplement_Fault_2014}, and it will be natural to see the connection between them. Recall that:

\begin{minipage}{0.4\textwidth}
\[
\Qcircuit @C=1.6em @R=1em {
 &\ctrl{1} &\qw \\
 &\targ &\qw
}\]
\end{minipage}
=
\begin{minipage}{0.4\textwidth}
\[
\Qcircuit @C=1.6em @R=1em {
&\qw &\ctrl{1} &\qw \\
&\gate{H} &\ctrl{0} &\gate{H}
}\]
\end{minipage}
\newline
\newline
Thus the circuit of Steane error correction in the $\hat{q}$ quadrature is equivalent to (neglecting the correction operator in \cref{fig:steane_circuit}):

\begin{minipage}{0.5\textwidth}
\[
\Qcircuit @C=1.6em @R=1em {
\lstick{ \ket{ \phi } }  & \ctrl{1} & \qw
\\
\lstick{ \ket{ + } }  & \targ  &\measureD{q_{out}}  
}\]
\end{minipage}
=
\begin{minipage}{0.5\textwidth}
\[
\Qcircuit @C=1.6em @R=1em {
\lstick{ \ket{  \phi }  }   & \ctrl{1} & \qw
\\
\lstick{ H\ket{ + } }      & \ctrl{0}  &\measureD{p_{out}}  
}\]
\end{minipage}
\newline
\newline
where measurement in the $\hat{q}$ quadrature after a Hadamard gate is equivalent to a measurement in $\hat{p}$ quadrature. Thus it's easy to see that this circuit is just two qubits connected by a CPHASE gate and measure the ancilla qubit prepared in state $\ket{0} = H \ket{+}$ in the $\hat{p}$ quadrature. Similarly, Steane error correction in the $\hat{p}$ quadrature is equivalent to:

\begin{minipage}{0.5\textwidth}
\[
\Qcircuit @C=1.6em @R=1em {
\lstick{ \ket{  \phi } }  &\targ  &\qw
\\
\lstick{ \ket{ 0 }}   &\ctrl{-1}  &\measureD{p_{out}}  
}\]
\end{minipage}
=
\begin{minipage}{0.5\textwidth}
\[
\Qcircuit @C=1.6em @R=1em {
\lstick{ H\ket{ \phi }  }  &\ctrl{0}  &\gate{H}
\\
\lstick{ \ket{ 0 } }  &\ctrl{-1}  &\measureD{p_{out}}  
}\]
\end{minipage}

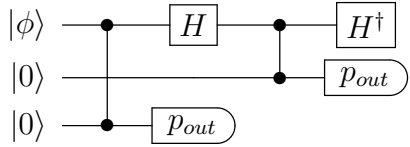
\begin{figure}
\centering
\begin{minipage}{\textwidth}
\[
\Qcircuit @C=1.3em @R=0.4em {
\lstick{ \ket{  \phi } }  & \ctrl{2} & \gate{H} &\ctrl{1} &\gate{H^{\dag}} 
\\
\lstick{ \ket{ 0 } }     &\qw       &\qw       &\ctrl{0} &\measureD{p_{out}} 
\\
\lstick{ \ket{ 0 } }     &\ctrl{0} &\measureD{p_{out}}
}\]
\end{minipage}
\caption{Circuit of quantum error correction consists of Steane error corrections in two quadratures. This circuit is equivalent to that in \cref{fig:steane_circuit} up to correction operators. All three qubits are connected by CPHASE gates, and Hadamard gates is quite easy to be implemented through cluster state. It's quite obvious to notice that Steane error correction fits nicely within the cluster state formalism.
}
\label{fig:steane_circuit_cphase}
\end{figure}

Similarly, there are also two qubits connected by a CPHASE gate. Neglecting the displacement operators, the whole circuit of Steane error correction is now written in \cref{fig:steane_circuit_cphase}.

\begin{figure}
\centering
\includegraphics[width=0.2\textwidth]{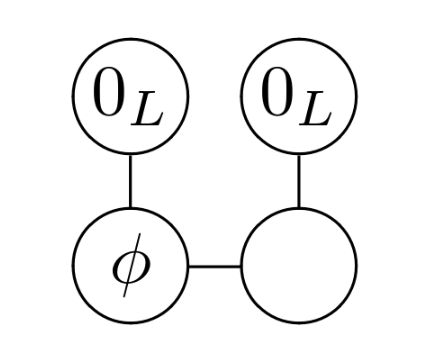}
\caption{The blank node represents a $\hat{p}$-squeezed vacuum state, i.e. $\ket{+}$ state. All links are CPHASE gate $\hat{C}_z = e^{ \mathrm{i}\hat{q}\otimes\hat{q}}$. After measuring each of the three marked qubits in the $\hat{p}$ quadrature, the blank node is left in state $\ket{\phi}$ with both quadratures corrected, up to a Hadamard.
}
\label{fig:cluster-Steane}
\end{figure}

In order to apply a Hadamard gate on the input qubit $\ket{\psi}$, we use a CPHASE gate to teleport the input state into the ancilla with a Hadamard gate applied:
\[
\Qcircuit @C=1.6em @R=1em {
\lstick{ \ket{  \phi }  }  &\ctrl{1}  &\measureD{p_{out}}
\\
\lstick{ \ket{ + } }  &\ctrl{0}  &\qw  
}\]
we write the data qubit in state $\ket{ \phi} = a \ket{0} + b \ket{1}$ with $\vert a \vert^2 + \vert b \vert^2 = 1$. It's easy to check that the CPHASE gate transforms the initial state $\ket{ \phi}\ket{+}$ into:
\begin{align*}
\ket{ \phi} \ket{+} & = ( a \ket{0} + b \ket{1}) \ket{+} \\
   &\to \frac{1}{\sqrt{2}} \ket{+} \big(  a \ket{+} + b \ket{-} \big)          +  \frac{1}{\sqrt{2}} \ket{-} \big(  a \ket{+} - b \ket{-} \big)
\end{align*}
when the measurement outcome on input qubit $\ket{\phi}$ produces eigenvalue $+1$,  the qubit collapses into state $\ket{+}$ and the ancilla was left in state $H\ket{ \phi} = a \ket{+} + b \ket{-}$. Otherwise there's an additional logical $\overline{Z}$ error on the ancilla. Since we always know whether there's a logical error, we can just correct the ancilla into state $H\ket{ \phi}$. Then we can use a CPHASE gate to connect this output ancilla with another fresh ancilla prepared in $\ket{0}$ and thus realize Steane error correction in the $\hat{p}$ quadrature. Up to Hadamard gates and outcome-dependent displacements, the whole circuit can be recognized as a standard continuous variable cluster state as in \cref{fig:cluster-Steane}. 

\chapter{Quantum Error Correction with continuous information}
\label{sec:QEC-continuous}

In this chapter, the continuous nature of the GKP code is taken into account. In Steane error correction, continuous shift errors mean that the homodyne measurement can give us more than just binary values. It will be clear that this observation can increase the fault-tolerance of the GKP code.

Section \ref{sec:further_steane} is a further analysis of Steane error correction scheme. Based on the continuous measurement outcomes, the shift errors of the output qubits are not simply replaced by those of ancillas. Also the output qubits would have conditional error rates depending on the measurement outcomes.

In Section \ref{sec:multiple_measure}, we propose a modified version of Steane error correction scheme with two measurements. Since the measurement outcomes are continuous, it's natural to try multiple measurements and correct the errors based on all these measurements instead of only the first one. There's indeed some improvement, but unfortunately it's nearly negligible for the error model in this thesis.

Section \ref{sec:bit_flip_code} concatenates the three-qubit bit flip code with the GKP code. As analyzed in Section \ref{sec:further_steane}, instead of only an average error rate, the underlying GKP encoded qubits now have varying error rates after Steane error correction, which enables us to do a maximum-likelihood decision instead of always choosing the cases of one-bit flip error. 

In Section \ref{sec:toric_gkp}, the toric code is concatenated with the GKP code. Similarly the conditional error rates after Steane error correction can be used to modify the decoding scheme,
thus achieve the error threshold with much noisier GKP code states. Section \ref{sec:decoding_toric} is a short introduction of the toric code and its decoding scheme based on minimum-weight matching algorithm. Section \ref{sec:toric-onlyData} assumes that only data are noisy, while the ancillas used in Steane error correction are perfect. With our proposed method, we can use much noisier GKP encoded qubits ( average error rate around $14\%$) to achieve the error threshold (10.3\%). In Section \ref{sec:toric-allGKP}, all qubits are noisy, which makes the syndrome measurements also noisy, and we propose a scheme to correct the defects conditionally before decoding the toric code. 

\cref{sec:MLD} is a short discussion about Maximum-Likelihood decoding, which incorporates the continuous information naturally. Finally, Section \ref{sec:gkp_discuss} is a summary of the whole chapter.

\newpage
\section{Further Analysis of Steane Error Correction}
\label{sec:further_steane}

Finitely squeezed GKP code states are composed of Gaussians weighted by a Gaussian envelope as discussed in \cref{sec:encoding scheme} \cite{gottesman2001encoding}. Zero variance means that Guassians goes back to delta functions, i.e. finitely squeezed states go back to ideal states. One can find the representation of finitely squeezed state $\ket{\overline{0}}$ in \cref{eq:approximate_states}.

In the following analysis we consider ideal GKP code states subjected to independent stochastic  Gaussian shift errors in both quadratures, i.e. errors from the Gaussian shift error channel in \cref{sec:Gaussian_shift_error} . For simplicity we write a GKP code state subjected to Gaussian shifts as (up to phase factors):
\begin{align*}
\ket{\psi, u, v} = e^{-iu\hat{p}}e^{-iv\hat{q}}\ket{ \overline{\psi} },
\end{align*}
where $\ket{ \overline{\psi} }$ represents an arbitrary ideal GKP code state. $u$, $v$ are independent Gaussian variables with variance $\sigma$, i.e. $P_{\sigma}(u) = \frac{1}{\sqrt{ 2\pi \sigma^2} } e^{-\frac{u^2}{2 \sigma^2}}$. We only consider the shifts in the $\hat{q}$ quadrature, and the analysis in the $\hat{p}$ quadrature is the same, so the state would be simplified as:
\begin{align*}
\ket{\psi, u} = e^{-iu\hat{p}} \ket{ \overline{\psi} },
\end{align*}

In \cref{sec:steane_qec}, we have already considered the success rate of Steane error correction \cite{terhal2016encoding} \cite{glancy2006error}. But what we obtained is only an average rate, because what we do is simply mapping the GKP code to a normal qubit. However, with different values of the homodyne measurement outcomes $q_{out}$, the situations should be different. In this section, we take this additional information into account, it will be clear that the logical error rates depend on the measurement outcomes.

\subsection{Conditional Output Errors }
\label{sec:cond_error_steane}

After Steane error correction in the $\hat{q}$ quadrature as shown in \cref{fig:Steane-QEC-q}, the shift error has been replaced by the ancilla's shift error $u_2$ plus a possible $\overline{X}$ error. But the measurement gives us more information about $u_1$ and $u_2$, thus we should update the probability distributions of them according to the Bayes' theorem.

Once we make the measurement and get the outcome $q_{out}$, which means that $u_1 + u_2 = q_{out} + n \sqrt{\pi}, n \in \mathbb{Z}$. From Bayes' theorem $P(A|B) = \frac{P(B|A)P(A)}{P(B)}$, we can calculate the conditional probability distribution of $u_1$ 
with respect to the measurement outcome $q_{out}$:
\begin{equation}
\begin{split}
\mathbb{P}\left(u_1|u_1+u_2=q_{out}+n\sqrt{\pi}\right)=&\frac{ \mathbb{P}(u_1+u_2=q_{out}+n\sqrt{\pi} \vert u_1)P(u_1)}{P(u_1+u_2=q_{out}+n\sqrt{\pi})}\\
=&\frac{P(u_2=q_{out}+n\sqrt{\pi}-u_1)P(u_1)}{P(u_1+u_2=q_{out}+n\sqrt{\pi})}.
\end{split}
\label{eq:cond_output_error}
\end{equation}
First, we know that $u_1$ and $u_2$ are independent Gaussian variables with variance $\sigma_1$ and $\sigma_2$ respectively. Then $u_1+u_2$ is also a Gaussian variable with variance $\sqrt{\sigma_1^2 + \sigma_2^2}$. Taking these probability distributions into the Eq.(\ref{eq:cond_output_error}), we get:
\begin{equation}
\begin{split}
\mathbb{P} \left(u_1|u_1+u_2=q_{out}+n\sqrt{\pi}\right)&=\frac{\frac{1}{\sigma_2 \sqrt{2\pi}}e^{-\frac{\left(u_1-q'-l\cdot\sqrt{\pi}\right)^2}{2\sigma_2^2}}\frac{1}{\sigma_1 \sqrt{2\pi}}e^{-\frac{u_1^2}{2\sigma_1^2}}}{\frac{1}{\sqrt{\sigma_1^2+\sigma_2^2} \sqrt{2\pi}}e^{-\frac{\left(q_{out}+n\sqrt{\pi}\right)^2}{2\left(\sigma_1^2+\sigma_2^2\right)}}}\\
&=\frac{1}{\sigma \sqrt{2\pi}}e^{-\frac{\left(u_1-u_c\right)^2}{2\sigma^2}}
\end{split}
\label{eq:sum_gaussian}
\end{equation}
where $\sigma^2 = \frac{\sigma_1^2 \sigma_2^2}{\sigma_1^2 + \sigma_2^2}$ and $u_c =\frac{\sigma_1^2}{\sigma_1^2+\sigma_2^2} \left( q_{out}+ n\sqrt{\pi} \right)$. So after the measurement, $u_1$ now satisfies a Gaussian distribution with variance $\sigma$ and mean value $u_c$. The probability distribution of $u_2$ is quite similar and we write them as:
\begin{equation}
\begin{split}
u_1 \sim \mathcal{N}(\frac{\sigma_{1}^{2}}{\sigma_{1}^{2}+\sigma_{2}^{2}}(q_{out}+n\sqrt{\pi}),\sigma)\\
u_2 \sim \mathcal{N}(\frac{\sigma_{2}^{2}}{\sigma_{1}^{2}+\sigma_{2}^{2}}(q_{out}+n\sqrt{\pi}),\sigma)
\end{split}
  \label{eq:condition_output}
\end{equation}
where $\mathcal{N}(x,y)$ represents a Gaussian distribution with mean value $x$ and variance $y$. It's easy to see that $u_1$ obtains an additional mean value $u_c =\frac{\sigma_1^2}{\sigma_1^2+\sigma_2^2} \left( q_{out}+ n\sqrt{\pi} \right)$ and the variance $\sigma$ is less than $\sigma_1$. 

Since we have updated the probability distributions of $u_1,u_2$ after the homodyne measurement, we cannot simply apply the correction operation as before as in \cref{fig:steane_circuit}. Considering that the exact value of $n$ is unknown, we can only correct the part $\frac{\sigma_1^2}{\sigma_1^2+\sigma_2^2} q_{out}$ in the $\hat{q}$ quadrature. Then it's equivalent to say that the output qubit is left with a Gaussian shift error $u_1 \in \mathcal{N}(0,\sigma)$ plus an additional constant shift error $\frac{\sigma_1^2}{\sigma_1^2+\sigma_2^2} n\sqrt{\pi}$.  

Strictly speaking, the constant shift error $\frac{\sigma_1^2}{\sigma_1^2+\sigma_2^2} n\sqrt{\pi}$ is not an integer multiple of $\sqrt{\pi}$, but it is always very close to a logical shift or a stabilizer, since we always assume that input data qubits are much noisier than the ancilla qubit, i.e $\sigma_2 \ll \sigma_1$. 

For simplicity, in the following analysis we assume the limit that $\frac{\sigma_2}{\sigma_1} \rightarrow 0$, and neglect the changes on the probability densities of $u_1$ and $u_2$. The analysis above is then reduced back to the original analysis of Steane error correction, i.e input shift error will be replaced by that of ancilla qubit after the correction step, and there's a logical error with some probability.

\subsection{Conditional Error Rates}
\label{sec:cond_rate_steane}

\begin{figure}
\centering
\begin{minipage}{\textwidth}
\[
\Qcircuit @C=2.0em @R=1.4em {
\lstick{\ket{\psi, u_1} }  & \ctrl{1} &\ustick{u_1} \qw & \qw &\gate{\exp(i q_{cor} \hat{p})} &\rstick{\ket{\psi, u_1 - q_{cor} }} \qw\\
\lstick{\ket{+, u_2} }  & \targ & \qw &\ustick{u_1+u_2 }  \qw  &\measureD{q_{out}} \cwx[-1] \\\\
}\]
\end{minipage}
\caption{Circuit of Steane error correction for shift errors in $q$ quadrature. The logical CNOT gate moves the shift error $u_1$ of $\ket{\psi,u_1}$ to the ancilla qubit, which leads to the homodyne measurement outcome $q_{out} = u_1 + u_2 + n \sqrt{\pi}, n\in \mathbb{Z} $. With $q_{cor} = q_{out} \mod \sqrt{\pi}$ in the range $[-\frac{\sqrt{\pi}}{2}, \frac{\sqrt{\pi}}{2}]$, the correction operator $\exp{(i q_{cor} \hat{p_1})}$ is then applied.}
\label{fig:Steane-QEC-q}
\end{figure}
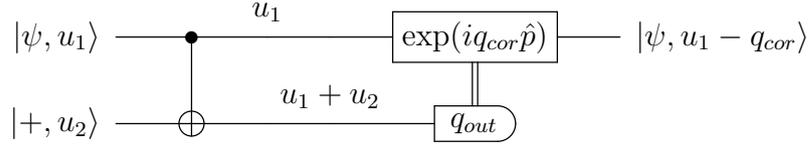

The circuit to correct shifts in the $\hat{q}$ quadrature is shown in \cref{fig:Steane-QEC-q}. As discussed in \cref{eq:steane_q_state_evolution}, the CNOT gate has the action:
\begin{align*}
\ket{\psi, u_1}_1\ket{+, u_2}_2 \to \ket{\psi, u_1}_1\ket{+, u_1 + u_2}_2.
\end{align*}
It's easy to see from \cref{eq:output_steane_q} that $q_{\rm out}$ should satisfy the relation:
\begin{align}
q_{\rm out} - u_1 - u_2 = n \sqrt{\pi},\quad n\in \mathbb{Z},
\label{eq:q_out}
\end{align}
which represents the tooth of the comb that one measures. Thus the probability density of $q_{\rm out}$ is:
\begin{equation}
\begin{split}
& \mathbb{P}(q_{\rm out}|u_1, u_2)= \frac{1}{\mathcal{N}}\sum_{n\in \mathbb{Z}} \delta(q_{\rm out} -u_1 - u_2 - n\sqrt{\pi} ),
\end{split}
\label{eq:conditional_q_out}
\end{equation}
where $\mathcal{N}$ is the normalization factor.

The correction $q_{cor}$ is defined as $q_{\rm cor} = q_{out} \mod \sqrt{\pi} $ in the range $ [-\frac{\sqrt{\pi}}{2}, \frac{\sqrt{\pi}}{2}]$ which, in case of $u_2=0$, would shift the input codeword plus error back to a codeword. If one finds a large $q_{\rm out}$ this could be due to large shifts, but it could also be due to just hitting a farther tooth in the comb (all teeth are equally likely). However, it is clear that if the found value for $q_{\rm cor}$ lies at the boundary of its interval, one is less certain about whether one has applied the right correction. We can also evaluate the logical $\overline{X}$ error probability given $q_{\rm out}$.

Assume that the data qubit and the ancilla qubit have shifts according to the Gaussian distribution $P_{\sigma_1}(u_1)$ and $P_{\sigma_2}(u_2)$. We have the probability density of $q_{\rm out}$:
\begin{align}
\mathbb{P}(q_{\rm out})&=\int du_1 \int du_2 P_{\sigma_2}(u_2)P_{\sigma_1}(u_1) 
\mathbb{P}(q_{\rm out}|u_1, u_2)\nonumber \\
&=\frac{1}{\mathcal{N}}\sum_{n \in \mathbb{Z}} \int du_1 P_{\sigma_1}(u_1)P_{\sigma_2}(q_{\rm out}-u_1-n\sqrt{\pi})
\label{eq:distribution_q_out}
\end{align}
where in principle $\int dq_{\rm out} \mathbb{P}(q_{\rm out})=1$.

When is the procedure succesful? When $|u_1+u_2-2k \sqrt{\pi}| < \sqrt{\pi}/2$ for some integer $k$ (that is, the shifts add up to a stabilizer shift plus less than half a logical shift), then the correction operator $ e^{i q_{cor}\cdot\hat{p_1}}$ will leave at most a remaining $u_2$ error. When $|u_1+u_2-2k \sqrt{\pi}| < \sqrt{\pi}/2$ we write that $u_1,u_2 \in I_{\rm succces}$. Further, We can write the conditional probability density of $u_1,u_2$ with respect to $q_{out}$ as:
\begin{equation}
\mathbb{P}(u_1,u_2|q_{\rm out})=\frac{\mathbb{P}(q_{\rm out}|u_1,u_2) P_{\sigma_1}(u_1) P_{\sigma_2}(u_2)}{\mathbb{P}(q_{\rm out})}.
\end{equation}
so that 
\begin{equation}
\mathbb{P}(\mathrm{succ}| q_{\rm out}) = \int_{I_{\rm success}} du_2  du_1 \mathbb{P}(u_1,u_2|q_{\rm out})
\label{eq:succ_q_out}
\end{equation}
and the average success probability is: 
\begin{equation}\label{eq:rate_error}
\mathbb{P}(\mathrm{succ}) = \int_{I_{\rm success}} du_2  du_1 P_{\sigma_1}(u_1) P_{\sigma_2}(u_2).
\end{equation}
Obviously the success region $I_{sucess}$ and the value of $q_{out}$ only depend on the value of $u_1+u_2$, we switch the integration over the variable $w=u_1+u_2$ for convenience. According to the summation rule of independent Gaussian variables, $w$ is a Gaussian variable with variance $\sigma = \sqrt{\sigma_1^2+\sigma_2^2}$. $I_{\rm success}$ corresponds to the constraint $|w-2k \sqrt{\pi}|< \sqrt{\pi}/2$. Then we can write $\mathbb{P}(u_1,u_2|q_{\rm out})$ with a different integration variable as:
\begin{equation}
\mathbb{P}(w|q_{\rm out})=P_{\sigma}(w) \frac{\sum_{n \in \mathbb{Z}} \delta(q_{\rm out}-w-n \sqrt{\pi})}{\sum_{n\in \mathbb{Z}} P_{\sigma}(q_{\rm out}-n\sqrt{\pi})}.\\
\end{equation}
So that: 
\begin{equation}\label{eq:out}
\begin{split}
\mathbb{P}(\mathrm{succ}| q_{\rm out})&= \int_{I_{\rm success}} \D w  \mathbb{P}(w|q_{\rm out})\\
& = \frac{1}{\sum_{n\in \mathbb{Z}} P_{\sigma}(q_{\rm out}-n\sqrt{\pi})} \int_{I_{\rm success}} \D w  P_{\sigma}(w) \sum_{n \in \mathbb{Z}} \delta(q_{\rm out}-w-n \sqrt{\pi})\\
&=\frac{\sum_{n \in \mathbb{Z}} P_{\sigma}(q_{\rm out}-n \sqrt{\pi})f_{\rm succes}(n,q_{\rm out}). 
}{\sum_{n\in \mathbb{Z}} P_{\sigma}(q_{\rm out}-n\sqrt{\pi})}.
\end{split}
\end{equation}
where $f_{\rm succes}(n,q_{\rm out})=1$ when $q_{\rm out}-n\sqrt{\pi}$ is at most $\sqrt{\pi}/2$ away from an even multiple of $\sqrt{\pi}$ and otherwise 0.
We note that the right hand side of the expression only depends on $q_{\rm out}-n \sqrt{\pi}$ and similarly $\mathbb{P}(q_{\rm out})$ is the same for any $q_{\rm out}$ plus multiple integers of $\sqrt{\pi}$. Hence we may restrict ourselves to considering a $q_{\rm out} \in [-\sqrt{\pi}/2,\sqrt{\pi}/2)$, which is the $q_{cor}$ defined ealier. 
This means that due to $f_{\rm success}$ the numerator on the r.h.s. of. Eq.(\ref{eq:out}) is restricted to even $n$, then the conditional success probability given $q_{out}$ can be written as :
\begin{equation}
\mathbb{P}(\mathrm{succ}| q_{\rm out}) = \frac{\sum_{n\in \mathbb{Z}} P_{\sigma}(q_{\rm out}-2n \sqrt{\pi})}{\sum_{n\in \mathbb{Z}} P_{\sigma}(q_{\rm out}-n\sqrt{\pi})}.
\end{equation}

Assume that $\sigma$ is small enoug, i.e. $\lvert w \vert = \lvert u_1+u_2 \rvert \leq \frac{2k+1}{2}\sqrt{\pi}$, with $k$ a positive integer. Since now $q_{out}$ has been restricted in the range $[-\sqrt{\pi}/2,\sqrt{\pi}/2]$, then we can make an approximation: 
\begin{equation}
\begin{split}
\mathbb{P}(q_{\rm out})&=\int \D w_1  P_{\sigma}(w) 
\mathbb{P}(q_{\rm out}|w) \\
&=\frac{1}{\mathcal{N}}\sum_{n \in \mathbb{Z}} \int dw_1 P_{\sigma}(w)\delta(q_{\rm out}-w-n\sqrt{\pi})\\
&\approx \frac{1}{\mathcal{N}_k}\sum_{|n|\leq k} P_{\sigma}(q_{\rm out} - n\sqrt{\pi}).
`\end{split}
\label{eq:distribution_q_out}
\end{equation}
where $\mathcal{N}_k$ is simply determined by the normalization of $\mathbb{P}(q_{\rm out})$ over $q_{\rm out}$ given that $\lvert w \vert \leq \frac{2k+1}{2}\sqrt{\pi}$. Similarly,  the success probability conditioned on $q_{\rm out}$ is approximately: 
\begin{equation}\label{eq:conditional_succ}
\mathbb{P}(\mathrm{succ}| q_{\rm out}) \approx \frac{\sum_{|2n| \leq k} P_{\sigma}(q_{\rm out}-2n \sqrt{\pi})}{\sum_{|n|\leq k} P_{\sigma}(q_{\rm out}-n\sqrt{\pi})}.
\end{equation}


\subsubsection{Post-Selection Based on Conditional Error Rates}
\label{sec:post-selection}
\begin{figure}
\centering
\includegraphics[width=0.7\textwidth]{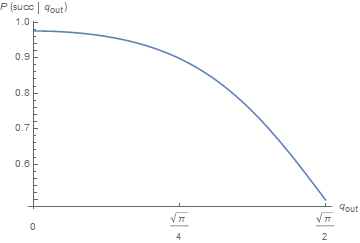}
\caption{Conditional success probability with respect to measurement outcome, with the variance of the shift error $\sigma = 0.6$. The x axis is the value of the homodyne measurement outcome $q_{out}$ , the y axis is the conditional success probability $\mathbb{P}(\mathrm{succ}| q_{\rm out})$, which is a monotonically decreasing function with respect to $q_{out}$ in range $[0, \sqrt{\pi}/2]$. Note that $\mathbb{P}(\mathrm{succ}| q_{\rm out})$ reaches its maximum when $ q_{\rm out} = 0$, which is around 0.975. }
\label{fig:cond_prob_0.6}
\end{figure}

Assume that the qubits are squeezed enough and shift errors are localized around 0, we set $k = 1$ so that $\lvert w_1 \vert \leq \frac{3}{2}\sqrt{\pi}$ in \cref{eq:conditional_succ}. The conditional success rate can be written as:
\begin{equation*}
\mathbb{P}(\mathrm{succ}| q_{\rm out}) \approx \frac{P_{\sigma}(q_{\rm out})}{ P_{\sigma}(q_{\rm out}-\sqrt{\pi}) + P_{\sigma}(q_{\rm out}) + P_{\sigma}(q_{\rm out} + \sqrt{\pi})}.
\end{equation*}
Since $\mathbb{P}(\mathrm{succ}| q_{\rm out})$ is symmetric with respect to $q_{out} = 0$, we restrict $q_{\rm out} \in [0, \sqrt{\pi}/2]$, then it's easy to check that $\mathbb{P}(\mathrm{succ}| q_{\rm out})$ is monotonically decreasing, see \cref{fig:cond_prob_0.6}.

Thus we're able to do a post-selection of qubits after Steane error correction. We can throw away any qubit $i$ with $q_{out,i} > q_{sel}$ where $q_{sel}$ is the selection criteria (an arbitrary number in the range $[0, \sqrt{\pi}/2]$), then with certainty the conditional success rates of the qubits remained are at least $\mathbb{P}(\mathrm{succ}| q_{sel})$. For the case in \cref{fig:cond_prob_0.6} with $\sigma = 0.6$, the maximum success rate in principle we can achieve is $\mathbb{P}(\mathrm{succ}| q_{sel} = 0 )  \approx 0.975$, comparing to the average success rate without post-selection which is only $0.86$.

Although his post-selection procedure is not practical for quantum computation, yet we can use it to lower the logical error rate while preparing a GKP-encoded qubit. For example, Fukui \textit{et. al} use this post-selection procedure to prepare cluster state\cite{FukuiGKPsurface2017}.

\subsubsection{Variance of Conditional Error Rates}

\begin{figure}
\centering
\includegraphics[width=0.7\textwidth]{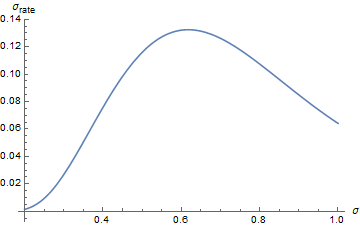}
\caption{$\sigma_{rate}$ is the variance of output error rates after Steane error corrections. $\sigma$ is the variance of the input qubits' shift errors}
\label{fig:variance}
\end{figure}

As in \cref{eq:conditional_succ}, the conditional success rate is a function of the measurement outcome $q_{out} \in [-\sqrt{\pi}/2,\sqrt{\pi}/2]$. Since this conditional success rate is itself a random variable satisfying a known probability distribution $\mathbb{P}(q_{\rm out})$  in \cref{eq:distribution_q_out}, there's also a variance of it, which we denote it as $\sigma_{rate}$ :
\begin{align*}
\sigma_{rate}^2 = \int_{-\sqrt{\pi}/2}^{\sqrt{\pi}/2}\mathrm{d}q_{out}\cdot\left( \mathbb{P}(\mathrm{succ}|q_{\rm out}) - \mathbb{P}(\mathrm{succ}) \right)^2\cdot \mathbb{P}(q_{\rm out})
\end{align*}
As shown in figure \ref{fig:variance}, $\sigma_{rate}$ is finite when the variance of input shift error $\sigma$ is around 0.5, which is non-trivial considering that $\sigma_{rate} =0$ with only an average error rate. This plot also fits our intuition: when $\sigma$, the variance of the input shift error, approaches zero, there's definitely no variance of the conditional error rate. When $\sigma$ approaches infinity, the probability distribution of the shift error will become a uniform random distribution over $[-\infty, \infty]$, then it will become completely random whether a measurement outcome $q_{out}$ is closer to an even or an odd multiple of $\sqrt{\pi}$, which also means $\sigma_{rate} = 0$ since the conditional probability is just a constant equal to $\frac{1}{2}$.

This non-trivial $\sigma_{rate}$ shows that the conditional success probability varies significantly. Thus some qubits are more likely to contain logical errors, which gives us a bias to modify our error correction and enables us to do maximum-likelihood decoding later in this thesis.


\section{Steane Error Correction with Multiple Measurements}
\label{sec:multiple_measure}

In last section, we see that the conditional error rates of Steane error correction depend on the homodyne measurement outcomes.  The conditional success rate $\mathbb{P}(\mathrm{succ} \vert  q_{out} )$ is monotonically decreasing with respect to $q_{out}  \in [0, \sqrt{\pi}/2]$. Then it's natural to wonder whether we could do more homodyne measurements to decrease the logical error rates. 

For example when we get $q_{out} = \sqrt{\pi}/2$, it's completely impossible to know which direction should we shift the state back, the correction will have half the probability to fail. Yet it might be possible to do another measurement before really shifting it back, and decide how we correct the error according to the information of two measurements.

Following this basic idea, we propose a modified version of Steane error correction: after two homodyne measurements, we decide how to correct the errors. Unfortunately for the error model in this thesis, it will be shown that double measurements in one Steane error correction only decrease the logical error rate trivially. Even worse, our proposed scheme introduces additional shift errors in the conjugate quadrature, given noisy ancilla qubits.

\subsection{Double measurements in One Steane Error Correction}
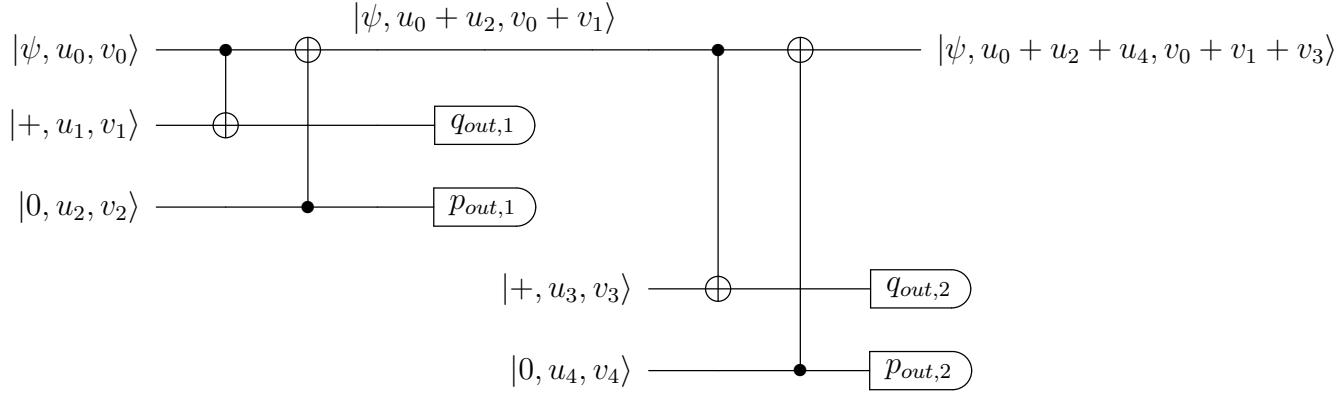
\begin{figure}
\centering
\begin{minipage}{0.7\textwidth}
\[
\Qcircuit @C=1.8em @R=1.4em {
\lstick{ \ket{\psi,u_0,v_0} } & \ctrl{1} & \targ &\qw&\ustick{\ket{\psi,u_0+u_2,v_0+v_1}}  \qw&\qw&\qw &\ctrl{3}  & \targ&\rstick{\ket{\psi,u_0+u_2+u_4,v_0+v_1+v_3} } \qw \\
\lstick{\ket{+,u_1,v_1}} & \targ &\qw &\qw &\measureD{q_{out,1}} \\
\lstick{\ket{0,u_2,v_2}} & \qw  &\ctrl{-2} &\qw&\measureD{p_{out,1}} \\
&&&&&&\lstick{\ket{+,u_3,v_3}}  &\targ &\qw&\measureD{q_{out,2}}\\
&&&&&&\lstick{\ket{0,u_4,v_4}} &\qw &\ctrl{-4} &\measureD{p_{out,2}}\\
 }\]
\end{minipage}
\caption{Circuit of the modified Steane error correction scheme without the correction steps. $u_0$ satisfy the Gaussian distribution $P_{\sigma_1}(u_0)$ with variance $\sigma_1$. $u_1,u_2,u_3,u_4$ satisfy the Gaussian distribution $P_{\sigma_2}(u)$ (Here we only focus on shift errors in the $\hat{q}$ quadrature). In order to prevent accumulating of shift errors, the measurements in the $\hat{q}$ and the $\hat{p}$ quadratures should be symmetric. Then this model is equivalent that we have noisy measurements, and after each measurement, there's some errors introduced. In principle this circuit could be easily generalized to more measurements.}
\end{figure}

First we do Steane error correction without really applying the correction operator, then the output qubit is in state $\ket{\psi,u_0+u_2,v_0+v_1}$, and from Eq.(\ref{eq:q_out}) the homodyne measurement outcome $q_{out,1}$ then is: 
\begin{equation}
\begin{split}
q_{out,1} = u_0 + u_1 + n_1 \sqrt{\pi} = w_1 + n_1 \sqrt{\pi}
\end{split}
\end{equation}
where $w_1 = u_0 + u_1$ with $n_1 \in \mathbb{Z}$. From Eq.(\ref{eq:conditional_q_out}), the probability distribution of $q_{out,1}$ conditioned on $w_1$ is:
\begin{equation}
\begin{split}
\mathbb{P}(q_{\rm out,1}|w_1) = \mathbb{P}(q_{\rm out,1}|u_0,u_1) &= \frac{1}{\mathcal{N}}\sum_{n_1\in \mathbb{Z}} \delta(q_{\rm out} -u_0 - u_1 - n_1\sqrt{\pi} )\\
&= \frac{1}{\mathcal{N}}\sum_{n_1\in \mathbb{Z}} \delta(q_{\rm out} -w_1 - n_1\sqrt{\pi} ).
\end{split}
\end{equation}
Then we can calculate the probability distribution of $q_{out,1}$:
\begin{equation}
\mathbb{P}(q_{\rm out,1})=\int du_0 \int du_1 P_{\sigma_1},(u_0)P_{\sigma_2}(u_1) 
\mathbb{P}(q_{\rm out}|u_0,u_1)
\end{equation}
when $|u_0+u_1-2k \sqrt{\pi}| < \sqrt{\pi}/2$ for some integer $k$ (that is, the shifts add up to a stabilizer shift plus less than half a logical shift), then the correction operator $ e^{-i q_{cor}\cdot\hat{p_1}}$ will leave at most a remaining $u_2$ error. When $|u_0+u_1-2k \sqrt{\pi}| < \sqrt{\pi}/2$ we write that $u_0,u_1 \in I_{\rm succces}$. Further, We can write 
\begin{eqnarray}
\mathbb{P}(u_0,u_1|q_{\rm out})=\frac{\mathbb{P}(q_{\rm out}|u_1,u_2) P_{\sigma_1}(u_1) P_{\sigma_2}(u_2)}{\mathbb{P}(q_{\rm out})}.
\end{eqnarray}
so that 
\begin{eqnarray}
\mathbb{P}(\mathrm{succ}| q_{\rm out,1}) &= \int_{I_{\rm success}} du_0  du_1 \mathbb{P}(u_0,u_1|q_{\rm out,1})
\end{eqnarray}
Given the measurement outcome $q_{\rm out,1}$, we know the conditional success probability if we really apply the correction operator of the Steane error correction. But we might be very unlucky to have $\lvert q_{out,1} \vert \approx \frac{2k+1}{2}\sqrt{\pi}  $ with some integer $k$, which will give us a very low success probability because we're not confident whether we can correct the data qubit in the right direction. Then it's natural to think whether we can do an additional Steane error correction to increase the success probability. As shown in the circuit of \cref{fig:double_measure}, we measure in the $\hat{q}$ quadrature once again to get $q_{out,2}$ with conditional probability distribution with respect to $w_2 =u_0 +u_2 + u_3$:
\begin{equation}
\begin{split}
\mathbb{P}(q_{\rm out,2}|w_2)&= \frac{1}{\mathcal{N}}\sum_{n_2\in \mathbb{Z}} \delta(q_{\rm out} -u_0 - u_2 -u_3 - n_2\sqrt{\pi} )\\
&=\frac{1}{\mathcal{N}}\sum_{n_2\in \mathbb{Z}} \delta(q_{\rm out} -w_2 - n_2\sqrt{\pi} ).
\end{split}
\end{equation}
Now the output qubit is in state $\ket{\psi,u_0+u_2+u_4,v_0+v_1+v_3}$, if now we apply the correction operator with $q_{out,2}$, then apart from a possible logical $\overline{X}$ error, the shift error in the $\hat{q}$ quadrature of the data qubit will be replaced by the sum of two ancillas' shift error $u_4 - u_3$ and we get output qubit in state $\ket{\psi,u_4-u_3,v_0+v_1+v_3}$. If we can determine the value of $w_2 = u_0 + u_2 + u_3$ correctly, then the whole correction procedure will succeed. So now we want to calculate the probability distribution $\mathbb{P}(w_2 \vert q_{out,1},q_{out,2})$ of $w_2$ given the values of $q_{out,1}$ and $q_{out,2}$:
\begin{equation}
\mathbb{P}(q_{out,1},q_{out,2}\vert w_1,w_2) = \frac{1}{\mathcal{N}^2} \sum_{n_1\in \mathbb{Z}}\sum_{n_2\in \mathbb{Z}} \delta(q_{\rm out,1} -w_1 - n_1\sqrt{\pi}) \delta(q_{\rm out,2} -w_2 - n_2\sqrt{\pi}).
\end{equation}
\noindent The probability distribution of obtaining $w_1$ and $w_2$ is :
\begin{equation}
\begin{split}
\mathbb{P}(w_1,w_2) &= \int du_0 P_{\sigma_1}(u_0) \prod\limits_{i=1}^3\int du_i P_{\sigma_2}(u_i) \delta(w_1 -u_0 -u_1) \delta(w_2 -u_0-u_2-u_3)\\
&=\mathcal{N} e^{ -
\frac{\sigma_1^2}{4\sigma_2^2(2\sigma_2^2+3\sigma_1^2)} \left[2(w_1-w_2)^2+6w_1^2+3w_2^2\right] ,
 }
\end{split}
\end{equation}
where two Gaussian integrations give us two constants, which are absorbed into the normalization factor $\mathcal{N}$. Now we can get the conditional probability distributions of $w_1,w_2$ given $q_{out,1}$ and $q_{out,2}$:
\begin{equation}
\begin{split}
&\mathbb{P}(w_1,w_2\vert q_{out,1}, q_{out,2}) = \frac{\mathbb{P}( q_{out,1}, q_{out,2}\vert w_1,w_2 ) P(w_1,w_2) } { \mathbb{P}( q_{out,1},q_{out,2}) }\\
&= \mathcal{N} \sum_{n_1\in \mathbb{Z}}\sum_{n_2\in \mathbb{Z}} \delta(q_{\rm out,1} -w_1 - n_1\sqrt{\pi}) \delta(q_{\rm out,2} -w_2 - n_2\sqrt{\pi})e^{ -
\frac{\sigma_1^2}{4\sigma_2^2(2\sigma_2^2+3\sigma_1^2)} \left[2(w_1-w_2)^2+6w_1^2+3w_2^2\right] 
 }\\
&=\mathcal{N} e^{
-\frac{\sigma_1^2}{4\sigma_2^2(2\sigma_2^2+3\sigma_1^2)} \left[2(q_{out,1}-q_{out,2} -(n_1-n_2)\sqrt{\pi} )^2+6(q_{out,1}-n_1\sqrt{\pi})^2+3(q_{out,2}-n_2\sqrt{\pi})^2\right] }
\end{split}
\label{eq:w1-w2-conditional}
\end{equation}
where the denominator $\mathbb{P}(q_{out,1},q_{out,2})$ is a constant for specific $q_{out,1},q_{out,2}$ and it is absorbed by the normalization factor $\mathcal{N}$. Since the $\mathbb{P}(w_1,w_2\vert q_{out,1}, q_{out,2})$ is determined by $n_1, n_2$:
\begin{align}
\mathbb{P}(n_1,n_2\vert q_{out,1}, q_{out,2}) = \mathbb{P}(w_1,w_2\vert q_{out,1}, q_{out,2})
\end{align}
We note that the right hand side of the Eq.(\ref{eq:w1-w2-conditional}) only depends on $q_{\rm out,1}-n_1 \sqrt{\pi}$ and $q_{\rm out,2}-n_2 \sqrt{\pi}$, similarly $\mathbb{P}(n_1,n_2\vert q_{out,1}, q_{out,2})$ is the same for any $q_{\rm out,1}$ and $q_{out,2}$ plus an integer multiple of $\sqrt{\pi}$. Hence we may restrict ourselves to considering $q_{\rm out,1}, q_{\rm out,2} \in [-\sqrt{\pi}/2,\sqrt{\pi}/2)$. While considering small shift errors that $\lvert w_1 \rvert , \lvert w_2 \rvert \leq \frac{2k+1}{2}\sqrt{\pi}$ with k some interger, it's restricted that $n_1,n_2 = 0,\pm 1, \pm 2,\cdots \pm k$ given that:
\begin{equation}
\begin{split}
& q_{out,1}= u_0 + u_1 + n_1\sqrt{\pi} = w_1 + n_1 \sqrt{\pi} \\
& q_{out,2}= u_0 + u_2 + u_3 + n_2 \sqrt{\pi} = w_2 + n_2 \sqrt{\pi}
\end{split}
\end{equation}

Since we will apply a correction operator according to $q_{out,2}$, it's easy to see that when $n_2$ is even, the output qubit is only left with a small shift error $u_4-u_3$ if we apply the correction operator $e^{iq_{out,2} \hat{p}}$, otherwise the correction fails because there will be an additional logical error. So what we need to do is to determine whether $n_2$ is even or odd and we don't care about the value of $n_1$. Hence we define two quantities here:
\begin{align*}
&\mathbb{P}_1 = \sum_{\vert 2n\vert \leq k } \sum_{\vert n_1\vert \leq k}\mathbb{P}(n_1,n_2=2n\vert q_{out,1}, q_{out,2}),\\
&\mathbb{P}_2 =  \sum_{\vert 2n+1 \vert \leq k} \sum_{\vert n_1\vert \leq k}\mathbb{P}(n_1,n_2=2n+1\vert q_{out,1}, q_{out,2}).
\end{align*}
When $\mathbb{P}_1 \geq \mathbb{P}_2$ we say that $n_2$ is even, an then we apply a correction operator $e^{iq_{out,2}\hat{p}}$, otherwise we say $n_2$ is odd and apply the correction operator $e^{i(q_{out,2}+\sqrt{\pi})\hat{p}}$.

\subsubsection{Numerical Simulation}
In our simulation, we consider small shift errors in the approximation $\lvert w_1 \rvert , \lvert w_2 \rvert \leq \frac{2k+1}{2}\sqrt{\pi}$, where we take $k = 1$. Then it's restricted that $n_1,n_2 = 0,\pm 1$. And we fix the value of $q_{out,1}$ in the range $[0,\sqrt{\pi}/2]$, since $q_{out,1}$ is symmetric in the range$[-\sqrt{\pi}/2,\sqrt{\pi}/2)$. 

$u_0$ and $u_1$ are independent Gaussian variables with variance $\sigma_1$ and $\sigma_2$ respectively, and $u_0 + u_1 = q_{out,1} + n_1 \sqrt{\pi}$. From Eq.\ref{eq:condition_output} we get the conditional probability distributions of $u_0$ and $u_1$ after the measurement:
\begin{align}
\label{eq:condition_u_0}
u_0 \sim \mathcal{N}(\frac{\sigma_{1}^{2}}{\sigma_{1}^{2}+\sigma_{2}^{2}}(q_{out}+n_1\sqrt{\pi}),\sigma)\\
u_1 \sim \mathcal{N}(\frac{\sigma_{2}^{2}}{\sigma_{1}^{2}+\sigma_{2}^{2}}(q_{out}+n_1\sqrt{\pi}),\sigma)
\end{align}
where $\sigma^2 = \frac{\sigma_1^2 \sigma_2^2}{\sigma_1^2 + \sigma_2^2}$. Note that $u_0$ and $u_1$ are now not independent, but it's not important because we don't care about $u_1$ at all since it will never appear again.

Given a fixed value of $q_{out,1}$, we can calculate the conditional probabilities of $n_1=-1, 0, +1$ respectively. Then for each round of simulation, we determine the number of $n_1$ according to its conditional probability distribution. And then randomly choose the value of $u_0$ according to the probability distribution in Eq.(\ref{eq:condition_u_0}). Since $u_2,u_3$ are also independent Gaussian variables with variance $\sigma_2$, we randomly choose the values of them. Then we get value of $w_2 = u_0 + u_1 + u_2$, which leads to the value of $q_{out,2} = w_2 \mod \sqrt{\pi}$ in the range $[-\sqrt{\pi}/2,\sqrt{\pi}/2)$. 

Given $q_{out,1}$ and $q_{out,2}$, we can calculate $\mathbb{P}_1$ and $\mathbb{P}_2$ to determine whether $n_2$ is even or odd. Finally we count how many times we succeed and then get the average logical error rate with double measurements and fixed $q_{out,1}$. 

For each $q_{out,1} \in [0,\frac{\sqrt{\pi}}{2})$, we can calculate the average error rate with a second measuremen, which is the red dashed line in the \cref{fig:double_measure}. The conditional error rate $\mathbb{P}(\rm succ \vert q_{out,1})$ with only one measurement is represented by the blue dashed line in Figure.(\ref{fig:double_measure}). 

\subsubsection{Discussion}
\begin{figure}
\begin{minipage}[t]{0.5\textwidth}
\centering
\includegraphics[width=1\textwidth]{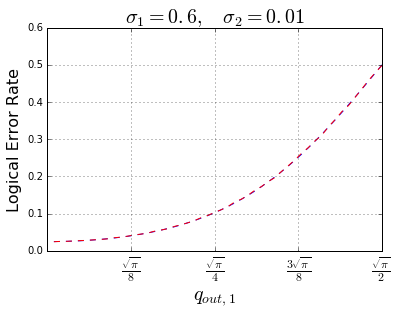}
\end{minipage}
\begin{minipage}[t]{0.5\textwidth}
\centering
\includegraphics[width=1\textwidth]{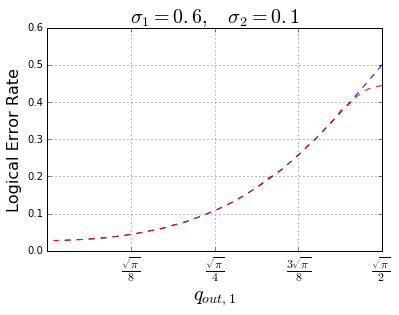}
\end{minipage}
\\
\begin{minipage}[t]{0.5\textwidth}
\centering
\includegraphics[width=1\textwidth]{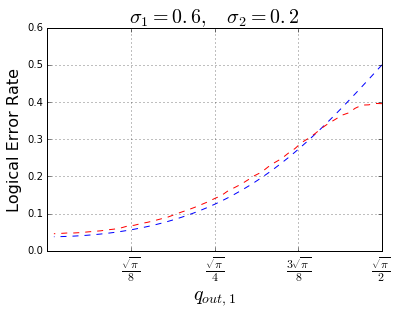}
\end{minipage}
\begin{minipage}[t]{0.5\textwidth}
\centering
\includegraphics[width=1\textwidth]{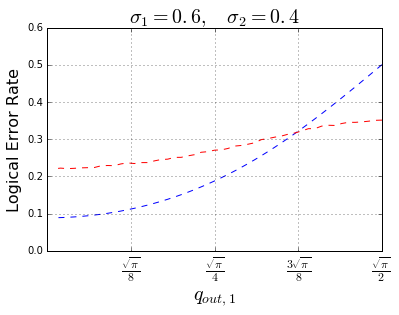}
\end{minipage}
\caption{The x axis is the first measurement outcome $q_{out,1}$, the y axis is the logical error rate. The blue dashed line represents the conditional error rate $\mathbb{P}(\overline{X} \vert q_{out,1})$ depending on $q_{out,1}$. With this fixed $q_{out,1}$, the red dashed line represents the average logical error rate if we do  a second measurement. It's clear that only when the first measurement outcome is quite close to $\frac{\sqrt{\pi}}{2}$, the second measurement can decrease the logical error rate.
}
\label{fig:double_measure}
\end{figure}

As shown in \cref{fig:double_measure}, the red dashed line is below the blue one when $q_{out,1}$ is quite close to $\frac{\sqrt{\pi}}{2}$. It gets worse when $q_{out,1}$ is small. Thus we need to decide whether we do the second measurement according to the $q_{out,1}$, making sure that the second measurement always decrease the logical error rate. 

However, considering that the probability to get $q_{out,1}$ close to $\frac{\sqrt{\pi} }{2}$ is very small, which means that it's not likely for a second measurement to make things better. We can estimate an upper bound of the improvement of error rate averaged over all $q_{out,1}$. In the case that $\sigma_1 = 0.6, \sigma_2=0.2$, here we do the second measurement only when $q_{out,1}$ is approximately larger than $0.4 \sqrt{\pi}$ with probability of about $5\%$, where the decrease can be optimistically estimated as a constant $5\%$. Then the total decrease of the error rate averaged over $q_{out,1} \in [0,\sqrt{\pi}/2]$ is less than $5\% \times 5\% = 0.25\%$, which is unfortunately negligible at all.

\section{Three-Qubit Bit-Flip Code with the GKP Code}
\label{sec:bit_flip_code}

The three qubit bit flip code is a very simple code that encodes a logical qubit into three physical ones and can detect and correct a single bit flip error\cite{nielsen2000quantum} \cite{PhysRevLett.119.180507}. In this section, we concatenate the three-qubit bit flip code with the GKP code and try to use the GKP error information (the conditional error rates) to do a maximum-likelihood decoding. Define the encoded qubit as:
\begin{align*}
\ket{\bar{0}} &= \ket{0}\ket{0}\ket{0},\\
\ket{\bar{1}} &= \ket{1}\ket{1}\ket{1}.
\end{align*}
While measuring the physical qubits would destroy the state of the system, it is possible to measure the parity between any two of them as the parity between two physical qubits contains no information about the logical state of the system.
A nice feature of these parity measurements is that they discretize the set of possible errors.
Let:
\begin{align*}
\ket{\tilde{0}} = \big(\sqrt{1 - p_E}\ket{0}+\sqrt{p_E}\ket{1}\big)\ket{0}\ket{0} .
\end{align*}
The parity check $Z_1 Z_2$ projects this state onto the code state $\ket{\bar{0}} = \ket{0}\ket{0}\ket{0}$ for the result $+1$ (with probability $(1 - p_E)$) and onto the ``error state'' $\ket{\tilde{0}} = \ket{1}\ket{0}\ket{0}$ for the result $-1$ (where $p_E$ is the average error rate for the three qubits).
In the next step, the qubit that got flipped is determined with a second parity check
$Z_2 Z_3$ and the error is corrected by applying an appropriate Pauli gate. The operators $Z_1 Z_2$ and $Z_2Z_3$ are the stabilizers of this code.

Note that this code only corrects bit flips, but not phase flips. However, the correction
of phase flips is completely analogous in the $\ket{+},\ket{-}$ basis, using $X_1 X_2$ and $X_2 X_3$ as parity measurements.

\subsection{Concatenation of the repetition code with the GKP Code}
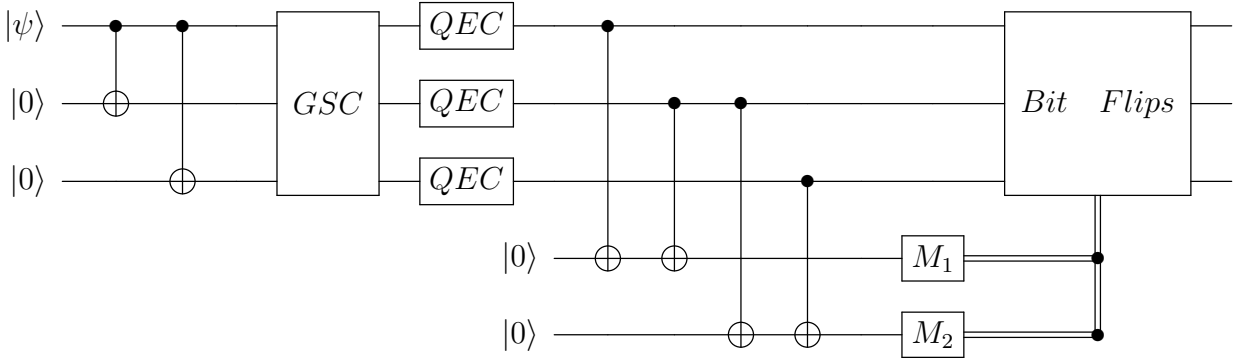
\begin{figure}
\centering
\begin{minipage}{\textwidth}
\[
\Qcircuit @C=1.3em @R=1em {
\lstick{ \ket{\psi} } & \ctrl{1} & \ctrl{2} &\qw &\multigate{2}{GSC}  &\gate{QEC} & \qw &\ctrl{3}  & \qw & \qw &\qw &\qw &\qw &\multigate{2}{Bit\quad Flips} &\qw \\
\lstick{\ket{0}} & \targ &\qw &\qw &\ghost{GSC}  &\gate{QEC} &\qw &\qw &\ctrl{2} &\ctrl{3} &\qw &\qw &\qw&\ghost{Bit \quad Flips} &\qw \\
\lstick{\ket{0}} & \qw  &\targ &\qw &\ghost{GSC} &\gate{QEC} &\qw  &\qw  &\qw &\qw &\ctrl{2} &\qw &\qw &\ghost{Bit \quad Flips} &\qw \\
&&&&&&\lstick{\ket{0}}  &\targ &\targ &\qw &\qw &\qw &\gate{M_1} &\cctrl{-1}\\
&&&&&&\lstick{\ket{0}}  &\qw &\qw &\targ &\targ &\qw &\gate{M_2} &\cctrl{-1}\\
 }\]
 \end{minipage}
\caption{Circuit of three-qubit bit flip code\cite{nielsen2000quantum} \cite{PhysRevLett.119.180507} concatenated with GKP codes. It's assumed that all GKP encoded qubits are prepared perfectly. After two CNOT gates to encode the repetition code, each qubit goes through a Gaussian shift error channel(GSC) and obtain independent gaussian shift errors in the $\hat{q}$ quadrature with variance $\sigma$, then we do Steane error correction with ideal ancillas. Finally we measure the stabilizer checks of the bit flip code and apply the correcting operations(bit flips) according to the measurement outcomes $M_1, M_2$ in the $\hat{q}$ quadrature, they corresponds to $Z_1Z_2$ and $Z_2Z_3$ respectively.}
\label{fig:bit-flip}
\end{figure}

For simplicity, we assume that all underlying GKP-encoded qubits are prepared perfectly. After two CNOT gates to encode the repetition code, they go through a Gaussian shift error channel(GSC) and obtains independent gaussian shift errors in the $\hat{q}$ quadrature with variance $\sigma$, see \cref{fig:bit-flip}. After Steane error correction with perfect ancillas, we know a conditional error rate for each GKP-encoded qubit. These rates are written as $p_1,p_2,p_3$.

It's easy to see that now we're able to make a maximum-likelihood decision between one bit flip error and double bit flip errors, write the probability for these two cases as $\mathbb{P}_1,\mathbb{P}_2$. The syndrome, $Z_1Z_2 = -1$  and $Z_2Z_3 = +1$ , for example, corresponds to two cases: (1) only the first qubit has a bit flip error; (2) only the first qubit has no error. We write the corresponding probabilities as:
\begin{align*}
&\mathbb{P}_1 = p_1\cdot (1-p_2) \cdot (1-p_3),\\
&\mathbb{P}_2 = (1-p_1)\cdot p_2 \cdot p_3.
\end{align*}
With only an average error rate, we will always find that $\mathbb{P}_1 > \mathbb{P}_2$ since we only consider small error rates. But with the GKP error information, $p_1,p_2,p_3$ are differentiated and it's possible that $\mathbb{P}_1 < \mathbb{P}_2$. Thus we're able to make a maximum-likelihood decision about the errors. The whole circuit is shown in Fig.(\ref{fig:bit-flip}).

\subsubsection{Numerical Simulation}
We use Monte Carlo method to do the simulation. First we assign each qubit with an independent Gaussian shift error with variance $\sigma$, and then simulate Steane error correction to get the conditional error rates for all the qubits, see Eq.(\ref{eq:conditional_succ}).  

We also know which qubits have bit flip errors, which leads to the syndrome $M_1,M_2$. With the conditional error rates, we can calculate the probabilities $\mathbb{P}_1,\mathbb{P}_2$ of two cases fitting the syndrome. Finally we make a maximum-likelihood decision to choose the case with larger $\mathbb{P}$.

\begin{figure}
\centering
\includegraphics[width=0.7\textwidth]{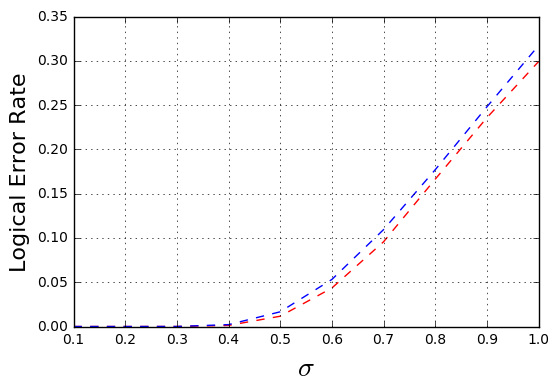}
\caption{The logical error rates of correcting three-qubit bit-flip code with or without using the conditional error rates of the underlying GKP-encoded qubits. Red dashed line represents correcting with the conditional error rates. The blue dashed line represents correcting with only the average error rate.  When the variance $\sigma$ of the input shift error is large enough ($\geq 0.4$), the logical error rates can be decreased a little bit.}
\label{fig:numerical_3qubits}
\end{figure}

The numerical results in Fig.(\ref{fig:numerical_3qubits}) shows that our proposed scheme can decrease the logical error rate non-trivially when the input shift errors are noisy enough, i.e variance $\sigma$ larger than 0.4. There's a similar scheme of three-qubit bit-flip code concatenated with the GKP code proposed by Fukui \textit{et al.} \cite{PhysRevLett.119.180507}, in which they didn't use Steane error correction.


\section{Concatenation of the toric Code with the GKP Code}
\label{sec:toric_gkp}

The toric code is defined as a square lattice with periodic boundary condition \cite{terhal2015quantum}. In this section, we consider an $L \times L$ two-dimensional toric code, which could be regarded as a torus, i.e. the right most edges are identified with the leftmost edges, and upper edges with lower edges. Each edge on the lattice is associated with a qubit and it is stabilized by plaquette operator $B_z = \prod_{j}Z_j$ and start operator $A_x = \prod_{j} X_j$ as shown in \cref{fig:toric}.

The error correction of two-dimensional toric code is a well-studied problem, including its decoding scheme as well as the error threshold\cite{criger2016noise}. In this section, we concatenate the toric code with the GKP code and try to use the GKP error information into account. It will be clear that we can achieve the error threshold with less squeezed GKP states.

\subsubsection{Decoding the Toric Code}
\label{sec:decoding_toric}

The error model we consider here is that qubits on each edge go through an error channel and get bit/phase flip errors with a constant probability $p_0$ independently , and the bit flip error is assumed to be independent from the phase flip errors. With possible bit/phase flip errors on the qubits, some stabilizers may produce $-1$ as outcome, such a stabilizer is called a defect.

\begin{figure}
\centering
\includegraphics[width=1\textwidth]{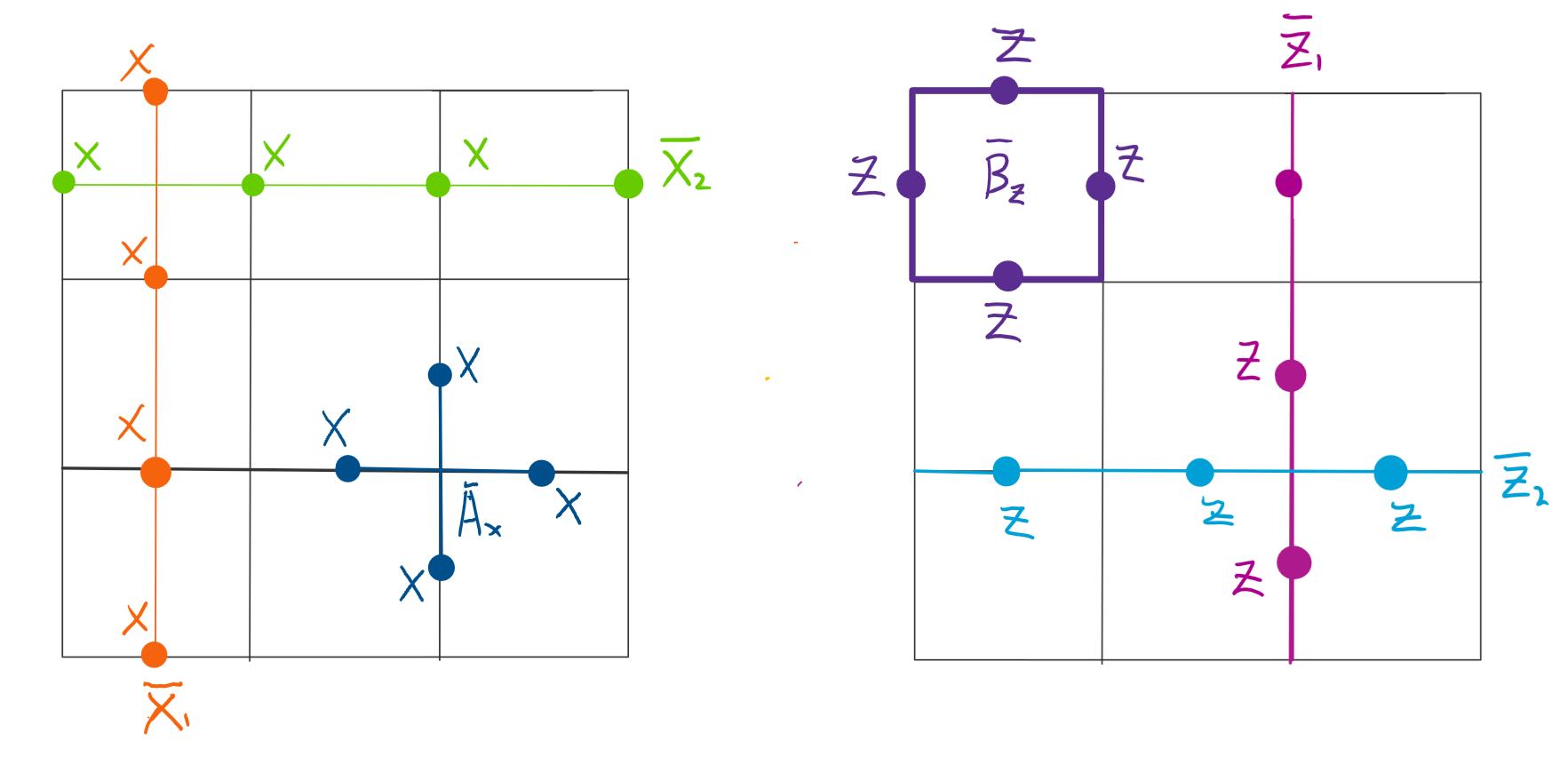}
\caption{Two dimensional toric code. Where $\overline{A}_x$ is the $X$ check, it consists of four $X$ operator acting on four qubits. $\overline{B}_z$ is the $Z$ check, it consists of four $Z$ operator acting on four qubits. $\overline{X}_i$ and $\overline{Z}_j$ are logical operators of the toric code, it's easy to check they commute both stabilizers, thus they are undetectable errors. $\overline{X}_i$ and $\overline{Z_j}$ don't commute with each other.($i,j=1,2$ )}
\label{fig:toric}
\end{figure}

Given a set of defects, we find paths with minimum sum of lengths to pair them up and bit/phase flip all the qubits on these paths, then these defects would disappear and the Toric code is again stabilized. But there's possibility to have a logical error which commutes with the stabilizers and is thus undetectable, see the logical errors $\overline{X}$ and $\overline{Z}$ in \cref{fig:toric}. The process described above is the well-known minimum-weight perfect-matching algorithm\cite{criger2016noise}. For the Toric code with only bit/phase flip errors on the data qubits, the theoretical error threshold is about $10.3\%$\cite{criger2016noise} \cite{gottesman2009introduction}, which means that under this threshold we could reach arbitrarily low logical error rate as we increase the size of the toric code.
\newline
\newline
\newline
Next we concatenate the toric code with the GKP code, replacing each qubit on the edge by a GKP-encoded qubit. Then we use the GKP error information of each qubit to modify the decoding process above, it will be shown that we can achieve the error threshold with noisier GKP code states.

\subsection{Only Data Qubits are Noisy}
\label{sec:toric-onlyData}

In this section, we assume that all GKP-encoded data qubits and ancilla qubits are prepared perfectly, but data qubits will go through the Gaussian shift error channel (see \cref{sec:Gaussian_shift_error}), and obtain Gaussian shifts. 

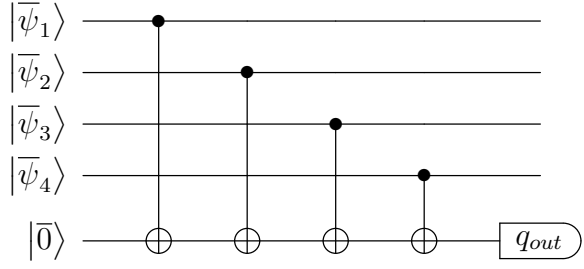
\begin{figure}
\centering
\begin{minipage}{\textwidth}
\[
\Qcircuit @C=2.0em @R=1.4em {
\lstick{\ket{ \overline{\psi}_1 } }  &\ctrl{4}  &\qw      &\qw       &\qw     &\qw  \\
\lstick{\ket{ \overline{\psi}_2 } }   & \qw      &\ctrl{3} &\qw       &\qw     &\qw \\
\lstick{\ket{ \overline{\psi}_3 } }   & \qw      &\qw      &\ctrl{2}  &\qw     &\qw         \\ 
\lstick{\ket{ \overline{\psi}_4 } }   & \qw      &\qw      &\qw       &\ctrl{1} & \qw           \\ 
\lstick{\ket{ \overline{0} } }       & \targ    &\targ    &\targ     &\targ &\measureD{q_{out}} \\ 
}\]
\end{minipage}
\caption{Circuit of stabilizer check $B_z = \prod_{j}\overline{Z}_j$. All the data qubits can only have possible logical $\overline{X}$ errors, and ancilla is also ideal in this circuit, so the homodyne measurement outcome $q_{out}$ can only be  $n\sqrt{\pi}$, $n \in \mathbb{Z}$.}
\label{fig:Steane-QEC}
\end{figure}

Before decoding the toric code, we first apply Steane error correction on all underlying GKP-encoded qubits. The shift error $u_1$ of a data qubit is assumed to be Gaussian with variance $\sigma_1$ and $u_2$ of ancilla qubit is set to be 0 in \cref{eq:distribution_q_out}) and \cref{eq:succ_q_out}). With the approximation $\lvert u_1 \rvert \leq \frac{2k+1}{2}\sqrt{\pi}$, we have:
\begin{subequations}
\begin{equation}
\mathbb{P}(q_{\rm out})=\int du_1 P_{\sigma_1}(u_1) \mathbb{P}(q_{\rm out}|u_1) \approx\frac{1}{\mathcal{N}_k}\sum_{|n|\leq k} P_{\sigma_1}(q_{\rm out}-n\sqrt{\pi})
\end{equation}

\begin{equation}
\mathbb{P}(\mathrm{succ}| q_{\rm out})= \int_{I_{\rm success}} du_1 \mathbb{P}(u_1|q_{\rm out}) \approx \frac{\sum_{|2n| \leq k} P_{\sigma_1}(q_{\rm out}-2n \sqrt{\pi})}{\sum_{|n|\leq k} P_{\sigma_1}(q_{\rm out}-n\sqrt{\pi})}
\end{equation}
\label{eq:P(q out)}
\end{subequations}
After Steane error correction, we do the syndrome measurements, i.e. the plaquette operator $B_z = \prod_{j}Z_j$ as in \cref{fig:toric}. The circuit of this operator is in \cref{fig:Steane-QEC}). The data qubits in the circuit are all perfect GKP states after Steane error corrections with perfect ancillas. Of course they also contain possible logical $\overline{X}$ errors with known conditional error rates, see \cref{sec:steane_qec}. It's easy to check that the effect of this circuit is :
\begin{align}
\prod_{i=1}^4\ket{\overline{\psi}_i } \cdot\ket{\overline{0}}\rightarrow (\frac{1+\prod_{i=1}^{4}\overline{Z}_i}{2} )\ket{\overline{\psi} }\cdot\ket{\overline{0}} + (\frac{1-\prod_{i=1}^{4}\overline{Z}_i}{2} )\ket{\overline{\psi} }\cdot\ket{\overline{1}},
\end{align}
where $\ket{\overline{\psi} }= \prod_{i=1}^4\ket{\overline{\psi}_i }$ and $\overline{Z}_i$ is a logical operator $\overline{Z}$ that acts on qubit $i$. If the measurement outcome implies that ancilla stays in state $\ket{0}$, the four qubits are stabilized by this plaquette operator. Otherwise this stabilizer produces eigenvalue -1, and gives us a defect. 

\subsubsection{Toric Code Decoding with Message Passing}
\label{sec:message-passing}

Instead of only an average logical success probability $\mathbb{P}({\rm succ})$ of the Steane error correction applied on the underlying GKP codes, one knows the conditional success probabilities depending on the measurement outcomes $q_{out}$. These conditional success probabilities for each qubit can be used in the minimum-weight-matching decoding process.

A path connecting any pair of defects in the toric code can be represented as a subset of qubits (or edges), and the probability for a subset $S$ in which all qubits have logical errors equals (here we write $\mathbb{P}(\overline{X}|q_{\rm out,i}) = 1- \mathbb{P}(succ|q_{\rm out,i})$ ):
\begin{align}
P_S = \prod_{i \notin S} \big(1-\mathbb{P}(\overline{X}|q_{\rm out,i}) \big) \prod_{j\in S} \mathbb{P}(\overline{X}|q_{\rm out,j}) = P_0 \prod_{j\in S} R_j
\label{eq:probability_subset}
\end{align}
where $R_j = \frac{\mathbb{P}(\overline{X}|q_{\rm out,j})}{1 -\mathbb{P}(\overline{X}|q_{\rm out,j})}$ and $P_0$ is the probability for no error on any underlying qubits of the toric code. For each pair of defects, now we need to find the path with largest $P_S$ instead of the path with shortest length. 

And considering that $P_0$ is the same for different subsets, the maximal $P_S$ means maximal $P_S/P_0 = \prod_{j\in S} R_j$, then we take a $\log$ function of it :
\begin{align}
\log{(\frac{P_S}{P_0})} = \log(\prod_{j\in S} R_j) = \sum_{j\in S} \log(R_j) = - \sum_{j\in S} w_j
\end{align}
where $w_j = - \log(R_j) = \log(\frac{1 -P(\overline{X}|q_{\rm out,j})}{P(\overline{X}|q_{\rm out,j})} )$. We assign each edge of the toric code with a weight equals to $w_j$, and then use the Dijkstra's algorithm for weighted graphs to find the path with largest $\log{(\frac{P_S}{P_0})}$, which corresponds to maximum $P_S$.

All the detected defects form a Graph $G$. For each pair of defects, we assign the maximum $P_S$ obtained above as the weight on the edge connecting them in $G$\cite{criger2016noise}. Then the minimum-weight-matching (Blossm) algorithm is run on $G$ to pair them up, thus determines which syndromes are matched. The Dijkstra's algorithm for weighted graph and minimum-weight matching (Blossom) algorithm are provided by a python library called networkx.

\subsubsection{Numerical Simulation}

To numerically simulate the decoding process, we randomly choose a shift error according to the Gaussian distribution with variance $\sigma_1$, resulting in a $q_{\rm out,i}$ for each qubit $i$. We imagine applying Steane error correction so that qubits in the toric code undergo an effective $\overline{X}$ error model with 
$\mathbb{P}(\overline{X}|q_{\rm out,i})=1-\mathbb{P}(\mathrm{succ}| q_{\rm out,i})$ for qubit $i$, with the average error rate $\mathbb{P}(\overline{X})$. We thus draw qubit errors for individual qubits from $P(\overline{X}|q_{\rm out,i})$. We decode the toric code for these errors and repeat the process of drawing $q_{\rm out}$ and drawing a qubit error to average over the variation of error rates.

\begin{figure}
\begin{minipage}[t]{0.5\textwidth}
\centering
\includegraphics[width=1\textwidth]{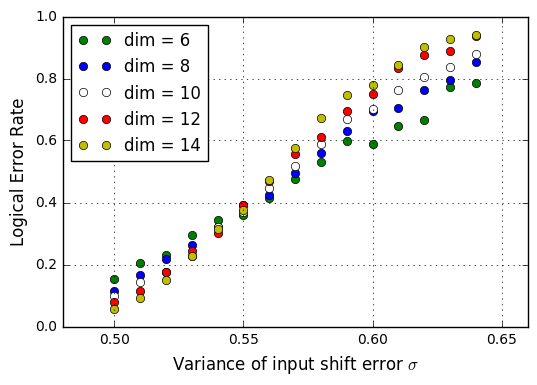}
\end{minipage}
\begin{minipage}[t]{0.5\textwidth}
\centering
\includegraphics[width=1\textwidth]{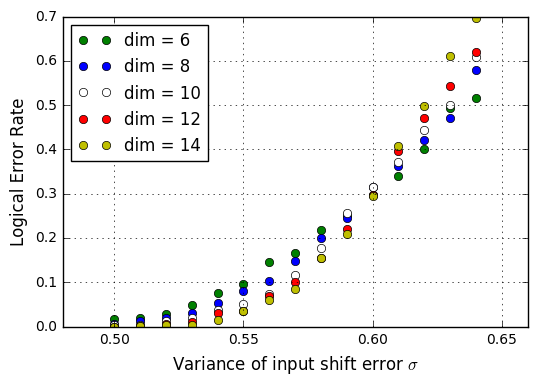}
\end{minipage}
\caption{Threshold comparison between decoding with or without continuous information. On the left, the simulation only takes the average error rate into account, and we achieve the error threshold between $\sigma \approx 0.54$ and $\sigma \approx 0.55$  corresponding to $\mathbb{P}(\overline{X}) \approx 10\%$ and $\mathbb{P}(\overline{X}) \approx 10.7\%$. The threshold fits nicely with the theoretical threshold of the toric code, $10.3\%$\cite{criger2016noise}.  On the right, the simulation takes the GKP error information (conditional error rates) into account, and the threshold is achieved with much noisier GKP states, where $\sigma \approx 0.6$ $(P(\overline{X}) \approx 14\%)$. "dim" in the legends means dimension of the 2-dimensional square toric code
}
\label{fig:toric-threshold}
\end{figure}

For each variance $\sigma$ there is an average $\overline{X}$ error rate $\mathbb{P}(\overline{X})= 1-\mathbb{P}(\mathrm{succ.})$ of a data qubit, see Eq.(\ref{eq:rate_error}). And the numerical results of the simulation is shown in Fig.\ref{fig:toric-threshold}. The left side is the simulation with only the average error rate, the right side uses the conditional error rates. 

With only an average error rate, each edge of the toric code is assigned with a constant weight, and the error threshold occurs between $\sigma = 0.54$ ($P(\overline{X}) \approx 10\%$) and $\sigma=0.55$ $P(\overline{X}) \approx 10.7\%)$. It fits nicely to the theoretical error threshold $10.3\%$ \cite{criger2016noise}. Using the GKP error information, each edge containing a qubit is assigned with a conditional error rate depending on $q_{out}$, then we can achieve the error threshold with an average error rate $\mathbb{P}(\overline{X}) \approx 14\%$ $(\sigma \approx 0.6)$. 

Though there's some inaccuracy in our simulation, the numerical results are already good enough to show that passing the conditional error rates to the minimum-weight matching decoder could tolerate much noisier GKP-encoded qubits. Furthermore, we can also calculate a conditional logical error rate of the toric code and use it in the next level of concatenation.

\subsection{Ancilla Qubits are Also Noisy}
\label{sec:toric-allGKP}

In this section, we assume that perfect data qubits and ancilla qubits all go through the Gaussian shift error channel. Shift error $u_d$ of data qubits and  $u_a$ of ancilla qubits are Gaussian variable with variance $\sigma_1$ and $\sigma_2$ respectively. For Steane error correction now we have:
\begin{subequations}
\begin{equation}
\begin{split}
\mathbb{P}(q_{\rm out})=\int dw P_{\sigma}(w) \mathbb{P}(q_{\rm out}|w)
\approx \frac{1}{\mathcal{N}_k}\sum_{|n|\leq k} P_{\sigma}(q_{\rm out}- w - n\sqrt{\pi}),
\end{split}
\end{equation}
\begin{equation}
\mathbb{P}(\mathrm{succ}| q_{\rm out})= \int_{I_{\rm success}} dw \mathbb{P}(w|q_{\rm out}) \approx \frac{\sum_{|2n| \leq k} P_{\sigma}(q_{\rm out}-2n \sqrt{\pi})}{\sum_{|n|\leq k} P_{\sigma}(q_{\rm out}-n\sqrt{\pi})},
\end{equation}
\label{eq:P(q out)}
\end{subequations}
where $w = u_a + u_d$ is a Gaussian variable with variance $\sigma = \sqrt{\sigma_1^2 + \sigma_2^2}$ and it's also approximated that $\lvert w \rvert \leq \frac{2k+1}{2}\sqrt{\pi}$ and restricted that $\lvert q_{out} \rvert \leq \sqrt{\pi}/2$. With $j=0,1$, the output state after correction is :
\begin{align}
\ket{\psi_{out}} = \ket{\psi, u_d - q_{cor} } = \exp{(-i (u_d-q_{cor})\cdot\hat{p_1})}\ket{ \overline{\psi} } = e^{i u_a\cdot\hat{p}} \overline{X}^j \ket{ \overline{\psi} }.
\end{align}
Note that $q_{cor} = u_a+u_d \mod \sqrt{\pi} \in [-\sqrt{\pi},\sqrt{\pi}]$, thus j equals 0 or 1.
\begin{figure}
\centering
\begin{minipage}{\textwidth}
\[
\Qcircuit @C=2.0em @R=1.4em {
\lstick{\ket{\psi_1,u_1}}  &\ctrl{4}  &\qw      &\qw       &\qw       &\ustick{u_1} \qw  &\qw        \\
\lstick{\ket{\psi_2,u_2}}   & \qw      &\ctrl{3} &\qw       &\qw      &\ustick{u_2} \qw  &\qw        \\
\lstick{\ket{\psi_3,u_3}}   & \qw      &\qw      &\ctrl{2}  &\qw      &\ustick{u_3} \qw  &\qw        \\ 
\lstick{\ket{\psi_4,u_4}}   & \qw      &\qw      &\qw       &\ctrl{1} &\ustick{u_4} \qw  &\qw         \\ 
\lstick{\ket{0,u_5}}        & \targ    &\targ    &\targ     &\targ    &\ustick{u_t}   \qw  &\measureD{q_{out}} \\ 
}\]
\end{minipage}
\caption{Circuit of stabilizer check $B_z = \prod_{j}\overline{Z}_j$. Where $u_t = \sum_{i=1}^5 u_i$ and its probability distribution is $P_{\sigma_t}(u_t)$ with $\sigma_t = \sqrt{5 \sigma_2^2}$, because all the data qubits and the ancilla qubit contain a Gaussian shift error $u_i$ with the same variance $\sigma_2$. Also all the data qubits have an possible logical $\overline{X}$ error. Then the homodyne measurement outcome $q_{out} = u_t + n\sqrt{\pi}$, $n \in \mathbb{Z}$. Note that $q_{out}$ later will also be restricted that $\lvert q_{out} \rvert \leq \sqrt{\pi}/2$.}
\label{fig:noisy_syndrome}
\end{figure}
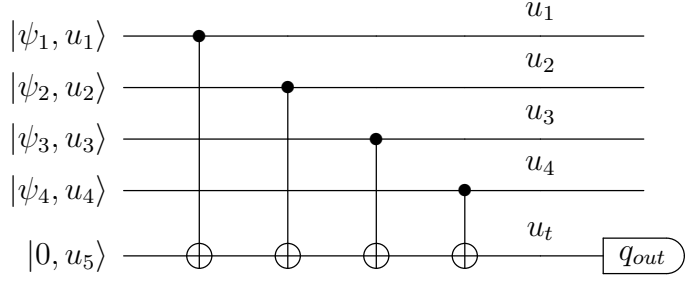

It's easy to see after the Steane error corrections, the shift error of the data qubit is replaced by ancilla's shift $u_a$ with a possible logical $\overline{X}$ error. After corrections we apply the syndrome measurements, the circuit of $B_z = \prod_{j}\overline{Z}_j$ is shown in \cref{fig:noisy_syndrome} and the effect of this circuit is:
\begin{align}
\ket{\psi,u} \cdot\ket{0,u_5}\rightarrow (\frac{1+\prod_{i=1}^{4} \overline{Z}_i}{2} )\ket{\psi,u}\cdot\ket{0,u_t} + (\frac{1-\prod_{i=1}^{4} \overline{Z}_i } {2} )\ket{\psi,u}\cdot\ket{1,u_t}
\label{eq:syndrome_measure_effect}
\end{align}
where $\ket{\psi,u} = \prod_{i=1}^4\ket{\psi_i,u_i}$ and $u_t = \sum_{i=1}^5 u_i$ is the ancilla qubit's total shift error after the CNOT gates. Given the homodyne measurement outcome $q_{out} =u_t + n\sqrt{\pi} $, we need to determine whether integer $n$ is even or odd. When $n$ is even, the ancilla is measured to be in state $\ket{0,u_t}$, which means $B_z$ produces an eigenvalue nearly $+1$. Otherwise the eigenvalue is nearly $-1$ and we have a defect. 

It's easy to see that we can make a right decision about $n$ to get a correct syndrome only when $u_t$ is closer to an even multiple of $\sqrt{\pi}$, i.e. $|u_t-2m \sqrt{\pi}| < \sqrt{\pi}/2$ for some integer m, which means the same successful region $I_{success}$ as the Steane error correction discussed in \cref{sec:cond_rate_steane}.

In order to deal with noisy syndrome measurements, we normally do multiple syndrome measurements\cite{fowler2012surface} \cite{criger2016noise}. For a $d \times d$ toric code, we do syndrome measurements $d$ times, then it's equivalent to decode a 3-dimensional($d \times d \times d$) toric code, and the error threshold is about $3\%$ when the qubit error rate is equal to the syndrome fail rate \cite{criger2016noise}. 

Now things are different, here the noisiness of syndrome measurements comes from noisy ancilla qubits, which is determined by $u_t = \sum_{i=1}^5 u_i$. Multiple syndrome measurements cannot give us too much information, because multiple measurements can only eliminate the effect of $u_5$ at most, we always have the noisiness of data qubits' remaining shift errors, i.e. $\sum_{i=1}^4 u_i$. Even worse the syndrome measurements will introduce shift error in the conjugate quadratrue (in practice we also need to deal with shift errors in the $\hat{p}$ quadrature), so we need to find another way to reduce the syndrome measurement error rate.

Fortunately, similar to the Steane error correction, the error rates of syndrome measurements also depend on the measurement outcome, which gives us a bias to achieve the goal, correcting the defects further before decoding the toric code.

\subsubsection{Correcting the Defects Conditionally}
\label{sec:post-selection}
 
Without loss of generality, in the following analysis we consider the case that $\sigma_1 = 2\sigma_2$, i.e. $\sigma_t = \sigma$. Thus we have $u_t = \sum_{i=1}^5 u_i$ and $w = u_a + u_d$ satisfy the same Gaussian distribution. Then it's easy to check that the syndrome measurements and Steane error corrections have completely the same average error rate $\mathbb{P}(\overline{X})$, also the same function of the conditional error rate $\mathbb{P}(\overline{X} \vert q_{out})$. 

Hence for each vertex, we use $P_i$ with $i = 1,2,3,4$ to denote the conditional $\overline{X}$ rates of four involved Steane error corrections, and $P_{syn}$ as the conditional fail rate of the syndrome measurement. With the five conditional error rates for each vertex, it's natural to think that some syndrome measurements are more likely to fail , and then we might be able to pick out and correct corresponding defects.

Provided ideal syndrome measurements that $P_{syn} = 0$, a stabilizer is detected as a defect only when one or three qubits contain errors, and the corresponding probability is:
\begin{align}
P_{defect} = \sum_{i=1}^{4}\left( (1-P_i)\cdot \prod_{i\ne j} P_j + P_i\cdot \prod_{i\ne j} (1-P_j) \right).
\end{align}
Now we consider noisy syndrome measurements with $P_{syn} > 0 $.  $P_{defect}\cdot (1 -P_{syn} )$ is probability that a defect's syndrome measurement succeeds and $(1-P_{defect})\cdot P_{syn}$ is the probability that a non-defect is detected to be a defect, i.e its syndrome measurement fails. 

The success rate of a syndrome measurement with eigenvalue $-1$ is:
\begin{align}
\mathbb{P}_{succ} = \frac{P_{defect}\cdot ( 1 - P_{syn} ) }{  P_{defect}\cdot (1-P_{syn})  + (1-P_{defect})\cdot P_{syn} }.
\label{eq:succ_defect}
\end{align}
With the average error rate $P(\overline{X})$ (remembering that Steane error correction and syndrome measurement have the same average error rate), it's quite straightforward to calculate the average success rate of the syndrome measurements with eigenvalue $-1$, we write it as $\bar{\mathbb{P}}_{succ}$ and it should satisfy:
\begin{align*}
\bar{\mathbb{P}}_{succ}= \lim_{N \to \infty} \frac{1}{N} \sum_{i = 1}^N  \mathbb{P}_{succ,i},
\end{align*}
where $N$ is the total number of vertices with eigenvalue $-1$, $\mathbb{P}_{succ,i}$ is the conditional success rate of the ith syndrome measurement with eigenvalue $-1$. The ratio of these defects with correct syndrome measurements is the average success rate written as: 
\begin{align}
\mathbb{R} = \bar{\mathbb{P}}_{succ} 
\end{align}
Note that we only consider syndrome measurements with eigenvalue $-1$, because the the success rate of eigenvalue $+1$ is nearly $100\%$ due to our assumption of small shift errors.

Of course, we're not satisfy with this ratio and want to increase it using the conditional error rates at hand. Based on the conditional success rate $\mathbb{P}_{succ}$ as in \cref{eq:succ_defect}, we propose a correcting process, which picks out part of vertices with large conditional success rate $\mathbb{P}_{succ,i}$, and regards the measurement outcomes of the rest as $+1$ in the following decoding process.

In the correcting process, we calculate the conditional success rate $\mathbb{P}_{succ,i}$ for each vertex $i$ with eigenvalue $-1$, and choose $P_c$ as a criteria to decide whether $\mathbb{P}_{succ,i}$ is large or small. Specifically, we compare $\mathbb{P}_{succ,i}$ with $P_c$ : if $\mathbb{P}_{succ,i} > P_c$ we say its syndrome measurement succeeds and it's indeed a defect, otherwise we correct this vertex to be a non-defect with eigenvalue $+1$. 

In order to maximize the ratio $\mathbb{R}$, we do simulations over $0 \leq P_c \leq 1$ and then pick the optimum $p_c$.

\subsubsection{Numerical Simulation}

\begin{figure}
\centering
\begin{minipage}{.7\linewidth}
\includegraphics[width=1\textwidth]{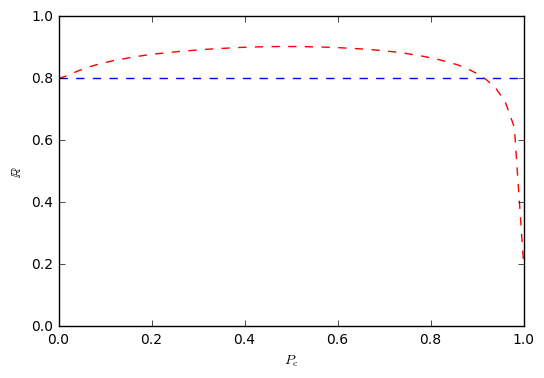}
\end{minipage}
\caption{The correcting process with $\sigma_1 = 2 \sigma_2 =0.4$, $\sigma = \sqrt{\sigma_1^2+\sigma_2^2} \approx 0.44$ and the average error rate $P(\overline{X}) \approx 4.75\%$. The $x$ axis is the the rate $P_c$ used as a criteria to decide whether a syndrome measurement is correct or not. Syndrome measurements will tell us that some vertices are defect, and the $y$ axis is the ratio $\mathbb{R}$ of the successful syndrome measurements, without the proposed correcting process it is around $80\%$ represented by the blue line and reaches the maximum when $P_c$ is around 0.5.}
\label{fig:post-selection}
\end{figure}

In the numerical simulation, we set $\sigma_1 = 2\sigma_2 = 0.4$ so that $\sigma = \sqrt{\sigma_1^2+\sigma_2^2} \approx 0.44$. The average error rate $P(\overline{X}) \approx 4.75\%$, with which it's easy to calculate the ratio of defects with correct syndrome measurements, i.e.  $\mathbb{R} = \bar{\mathbb{P}}_{succ} \approx 80\%$,  which is represented by the blue line as showed in Fig.(\ref{fig:post-selection}). The correcting process described above can increase the ratio as the red line in \cref{fig:post-selection}. Obviously $P_c \approx 0.5$ is the optimum $P_c$ to achieve the maximum $\mathbb{R} \approx 90\%$ in our simulation.

Note that $\mathbb{R} = 90\%$ corresponds to average error rate $P(\overline{X}) \approx 2\%$ if there's no such correcting process. $2\%$ means smaller variance of the shift error, i.e. $\sigma \approx 0.38 < 0.44$, thus the requirement of the noisiness of GKP code is relaxed.

\section{Beyond Minimum-Weight Matching}
\label{sec:MLD}

The decoding processes described above are all based on Minimum-weight perfect matching (MWM) algorithm. Even though we have used the continuous information to propose a modified version of it, this algorithm is itself not very good due to some intrinsic drawbacks\cite{bravyi2014efficient}

\subsection{Drawbacks of MWM}
\begin{figure}
\centering
\includegraphics[width=0.8\textwidth]{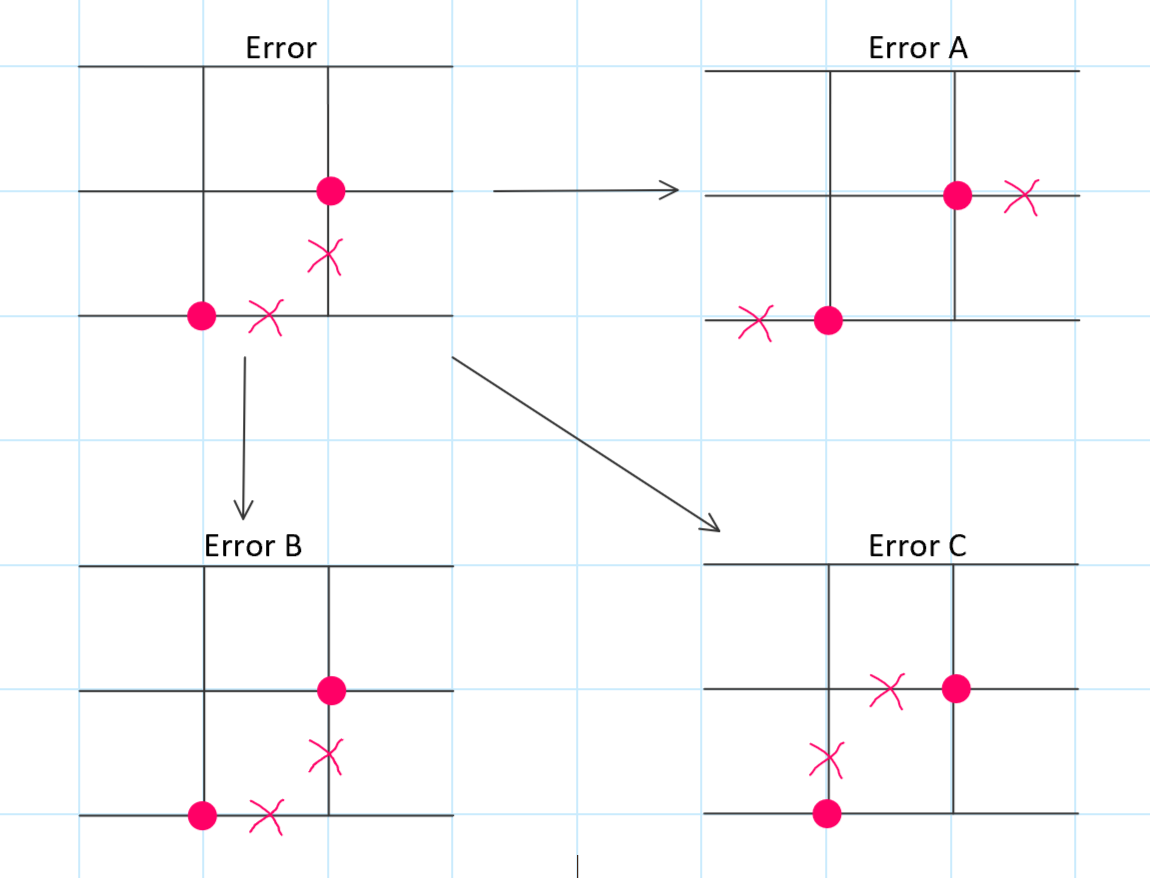}
\caption{Decoding process of Minimum-Weight Perfect Matching (MWM). Error A,B,C produces the same syndromes, which fits with the actual error. MWM cannot tell the difference between them ( continuous information can ameliorate this), and it's neither unable to account the equivalence of error B and C. Thus MWM is itself very likely to produce errors.}
\label{fig:drawback}
\end{figure}

In one word, the MWM decoder finds minimum-weight $\bar{X}$ or $\bar{Z}$ errors consistent with the observed syndromes and thus correct the errors. But this method is itself not good enough. First of all, minimum-weight matching does not mean minimum weight error, as shown in \cref{fig:drawback}, error A and B as well as C have the same syndromes, it's completely impossible to distinguish between them. As in \cref{sec:toric-onlyData}, with the varying error rates due to the continuous information, we're able to distinguish between them and thus ameliorate this drawback in some sense.

As shown in \cref{fig:drawback}, error  B, C only differ a stabilizer, they should be regarded as equivalent. Thus we should compare probability of error A with the sum of probabilities of error B and C, i.e. comparing $Pr(A)$ with $Pr(B) + Pr(C)$. However, MWM can only compare $Pr(A)$ with $Pr(B)$, or $Pr(A)$ with $Pr(B)$, which makes it easy to leave us a logical error. This problem of equivalence is considered in maximum-likelihood decoding algorithm.

\subsection{Maximum-Likelihood Decoder}
Considering these drawbacks, Dennis \textit{et al} \cite{dennis2002topological} \cite{bravyi2014efficient} introduced the Maximum-Likelihood Decoding (MLD) process. This decoder is based on a basic idea that any two operators that only differ a stabilizer is completely the same, i.e. actions on any encoded states are completely the same.

First we write $S$ as the group of stabilizer of the Toric Code. And $\bar{X}, \bar{Z}$ are the logical errors. Thus all the operators acting on the Toric code are divided into four equivalent classes: $S, \bar{X}S, \bar{Z}S, \bar{X}\bar{Z}S$, which means for any operator $g_1,g_2$ if they belong to the same class, for example $\bar{X}S$. Then we can choose either $g_1$ or $g_2$ as the actual error correct the Toric code. Using MLD, we fix some canonical error $E$ that fits the syndromes we observe, thus all the errors that fit the same syndromes are now divided into four equivalent classes:
\begin{align*}
 ES, \quad  E\bar{X}S, \quad E\bar{Z}S, \quad E\bar{X}\bar{Z}S
\end{align*}
For any group $G$ of these four classes, the probability of this group is defined as:
\begin{align*}
\mathbb{P}(G) = \sum\limits_{g \in G} P(g)
\end{align*}
where $P(g)$ is the probability that error $g$ occurs, where $g$ can be regarded as a subset of all the qubits in which qubits contain errors. From \cref{eq:probability_subset}, it can be written as:
\begin{align}
P(g) = \prod_i \big(1-P(\overline{X}|q_{\rm out,i}) \big) \prod_{j\in g} P(\overline{X}|q_{\rm out,j}) = P_0 \prod_{j\in g} R_j
\label{eq:probability_equivalence}
\end{align}
where $R_j = \mathbb{P}(\overline{X}|q_{\rm out,j})/\big(1 -\mathbb{P}(\overline{X}|q_{\rm out,j}) \big)$ and $P_0$ is the probability for no error. We decide that the errors fitting the observed syndromes belong to the equivalent class $G$ with largest $\mathbb{P}(G)$, and we choose any operator $g \in G$ to be the error and correct the toric code with respect to it.

Note that in \cref{eq:probability_equivalence}, we've already used the conditional error rates $\mathbb{P}(\overline{X}|q_{\rm out,i})$ of the GKP-encoded qubits, instead of only an average error rate as in the original proposal\cite{dennis2002topological}\cite{bravyi2014efficient}. The GKP error information thus fits nicely with the Maximum-Likelihood Decoder, using the conditional error rates to calculate the probabilities of the equivalent classes. Furthermore, this GKP error information can naturally be utilized in various decoding methods, like a Neural Decoder for Topological Codes \cite{torlai2017neural}.


\section{Discussion}
\label{sec:gkp_discuss}

In this chapter, we have taken the GKP error information into account. The quantum error correction protocol called Steane error correction is analyzed very carefully (see \cref{sec:further_steane}) and we proposed a modified version of it. (\cref{sec:multiple_measure}. Also we examined two error correction code concatenated with GKP code states ( \cref{sec:bit_flip_code} and \cref{sec:toric_gkp}). The numerical results shows that the GKP error information really relaxes the requirement of squeezing the GKP code.

As discussed in \cref{sec:cond_rate_steane}, based on the value of the homodyne measurement outcomes, it is possible to recognize qubits that are more likely to have logical errors. Following in \cref{sec:multiple_measure} we try to do multiple measurements in one Steane error correction, although the improvement is unfortunately negligible, but it at least give us some confidence to explore more in this direction.

Concatenating the three-qubit bit-flip code with GKP code states makes it possible to do maximum-likelihood decoding, if we take the GKP error information into account. 

In the data-only error model, the error threshold of the 2-dimensional toric code equals to $10.3\%$. For GKP-encoded qubits with stochastic Gaussian shift errors, this threshold corresponds to standard deviation $\sigma \approx 0.535$. However with our proposed decoding scheme, we can achieve this threshold with noisier GKP states with variance $\sigma \approx 0.6$, i.e. average error rate approximately $14\%$. Note that the proposed decoding method also fits naturally with the Maximum-Likelihood decoding as in \cref{sec:MLD}.

Further, various kinds of error correcting codes can be concatenated with the GKP code, for example the $C_4/C_6$ code discussed by Fukai \textit{et al} \cite{PhysRevLett.119.180507}. The information contained in the continuous shifts of the GKP code is taken into account, thus they improve the fault tolerance of the Bell measurements on which the $C_4/C_6$ code is based.

\begin{align*}
	 \rho \rightarrow (1-p)\rho + &p_x X\rho X + p_y Y\rho Y + p_z Z\rho Z \\
	& p = p_x + p_y + p_z \\
	& p_x = p_z\\
	& \eta = \frac{p_y}{p_x + p_z}\\
\end{align*}

Bias $\eta \rightarrow \infty$, only Y error.\\

If $\gcd(j,k) = 1$, then 

$j \times k$ surface code is equivalent to a classical repetition code.
\chapter{Conclusion and Outlook}
\label{sec:conclusion-outlook}
The continuous nature of the GKP code proves to be quite useful in quantum error correction, increasing fault tolerance of the GKP code. The basic idea of this thesis is quite simple: in error correction of GKP-encoded qubits, measurements in either quadrature collapse the qubits into different states, which depend on the measurement outcomes. This observation obviously gives us a bias to determine which qubits are more likely to contain errors.

Following this basic idea in \cref{sec:QEC-continuous}, we analyzed the conditional error rate of a qubit undergoing Steane error correction, to find that the error rate depends on the homodyne measurement outcome. This fact leads to many interesting results, where the most significant one is for the toric code. The conditional error rates of underlying GKP-encoded qubits can be used to modify the minimum-weight perfect-matching (MWM) of the toric code, achieving the error threshold with much noisier GKP code states. In some sense, our modified MWM is a special version of the maximum-likelihood decoding (MLD) algorithm, picking out the path with maximum probability instead of the one with shortest length. Note that this proposed method of using the GKP error information can be directly generalized into various error correcting codes via message passing. For example, the conditional error rates can be naturally implemented in MLD as discussed in \cref{sec:MLD}, and it remains to be answered how much we can improve MLD with the conditional error rates.

We mentioned a post-selection procedure of Steane error correction in \cref{sec:post-selection}. It can not be used in  quantum computation, but it should be useful in off-line preparation of the GKP-encoded qubits. It remains an open question whether this procedure can be incorporated in magic-state distillation for example.

In \cref{sec:MBQ_gkp} we discussed how Steane error correction fits nicely with cluster states. Menicucci \cite{Menicucci_Fault_2014} also analyzed the error bound of cluster states concatenated with GKP code. However, Menicucci didn't consider the conditional error rates and also forget that the measurements of cluster states are also noisy. Thus we need a further error analysis of the cluster states concatenated with the GKP code. It also remains an open question to explore: how to incorporate GKP error information into the framework of continuous variable measurement-based quantum computation.


\mainmatter



\appendix

\bibliographystyle{abbrv}
\bibliography{bibliography}
\end{document}